\documentclass[12pt]{article}%
\usepackage{graphicx}
\usepackage{amsmath}
\usepackage{amssymb}
\usepackage{authblk}
\usepackage{natbib}
\usepackage{setspace}
\usepackage{booktabs}
\usepackage{multirow}
\usepackage{hyperref}
\usepackage{lscape}
\usepackage{caption}
\usepackage{subfig}
\usepackage{rotating}
\usepackage{epsfig}
\usepackage{epstopdf}
\usepackage[hmargin=2.0cm,vmargin=2.0cm]{geometry}
\usepackage[margin=10pt,font=small,labelfont=bf]{caption}
\usepackage{array}
\usepackage{amsfonts}%
\usepackage{xcolor} 

\providecommand{\U}[1]{\protect\rule{.1in}{.1in}}
\hypersetup{
colorlinks=true,
linkcolor=black,
citecolor=blue,
filecolor=blue,
urlcolor=black,
}

\def\N{\mbox{N}}

\def\Q{\mbox{Q}}
\def\1{\mbox{1}}

\def\E{\mbox{E}}

\def\Y{\mbox{Y}}

\def\H{\mbox{H}}

\def\f{\mathbf{f}}

\def\bf{\mbox{\boldmath $\f$}}

\def\by{\mbox{\boldmath $y$}}

\def\bvarepsilon{\mbox{\boldmath $\varepsilon$}}
\def\bbeta{\mbox{\boldmath $\beta$}}

\def\N{\mbox{N}}

\def\Q{\mbox{Q}}
\def\1{\mbox{1}}

\def\E{\mbox{E}}

\def\Y{\mbox{Y}}

\def\f{\mbox{f}}

\def \U{\mbox{U}}

\def\H{\mbox{H}}

\def\f{\mathbf{f}}

\def\bvarepsilon{\mbox{\boldmath $\varepsilon$}}
\def\bbeta{\mbox{\boldmath $\beta$}}

\def\by{\mathbf{y}}

\def\bf{\mathbf{f}}

\def\bbeta{\mbox{\boldmath $\beta$}}

\def\bH{\mathbf{H}}

\def\bbeta{\mbox{\boldmath $\beta$}}

\def\bbeta{\mbox{\boldmath $\beta$}}

\def\bH{\mathbf{H}}

\def\by{\mathbf{y}}

\def\0{\mbox{\bf{0}}}

\def\E{\mbox{E}}

\def\N{\mbox{N}}

\def\by{\mathbf{y}}

\def\0{\mbox{\bf{0}}}

\def\by{\mathbf{y}}

\def\0{\mbox{\bf{0}}}

\def\N{\mbox{N}}

\def\E{\mbox{E}}

\def\H{\mbox{H}}
\def\Q{\mbox{Q}}

\def\N{\mbox{N}}

\def\H{\mbox{H}}

\def\Y{\mbox{Y}}

\def\N{\mathcal{N}}
\newcolumntype{C}[1]{>{\centering\let\newline\\\arraybackslash\hspace{0pt}}m{#1}}
\newcolumntype{R}[1]{>{\raggedleft\let\newline\\\arraybackslash\hspace{0pt}}m{#1}}

\usepackage{amsmath}
\usepackage{amsfonts}
\usepackage{amssymb}

\title{The Time-Varying Multivariate Autoregressive Index Model}
\author[1]{Gianluca Cubadda}
\author[1]{Stefano Grassi}
\author[2]{Barbara Guardabascio}
\affil[1]{University of Rome `Tor Vergata'}
\affil[2]{University of Perugia}

\begin{document}
\maketitle

\begin{abstract}
\noindent Many economic variables feature changes in their conditional mean
and volatility, and Time Varying Vector Autoregressive Models are often used
to handle such complexity in the data. Unfortunately, when the number of
series grows, they present increasing estimation and interpretation problems.
This paper tries to address this issue proposing a new Multivariate
Autoregressive Index model that features time varying means and
volatility.\newline Technically, we develop a new estimation methodology that
mix switching algorithms with the forgetting factors strategy of
\cite{KK2012}. This substantially reduces the computational burden and allows
to select or weight, in real time, the number of common components and other
features of the data using Dynamic Model Selection or Dynamic Model Averaging
without further computational cost.\newline Using USA macroeconomic data,
we provide a structural analysis and a forecasting exercise that demonstrates
the feasibility and usefulness of this new model.\\
\\
\textbf{Keywords}: Large datasets, Multivariate Autoregressive Index models,
Stochastic volatility, Bayesian VARs.

\end{abstract}

\newpage

\section{Introduction}

The availability of large datasets and instability of the economy has changed
the nature of economic models. Time-varying parameter models are developed to
capture the ever-changing economic environment. For example,
\cite{CS2002} use a small VAR with time-varying coefficients, that follow a random walk dynamic (TVP-VAR), to detect features such as coefficient drift of the
inflation-unemployment dynamics. Contributions on small TVP-VAR include
\cite{CMS2005}, \cite{CogleySargent(2005)}, \cite{Primiceri2005} and
\cite{dAgostinoGambettiGiannone2013}.\newline The abundance of large data sets
with macroeconomic variables has called for the development of larger TVP-VAR
models. Motivated by the need for modelling instability in large systems,
\cite{KK2013}(KK, henceforth) develop a computationally efficient estimation 
methodology for Large TVP-VAR models with stochastic volatility (TVP-VAR-SV). 
When a large number of predictors is included in the VAR
system, the computational burden increases significantly. KK operationalize
the model by using forgetting factors (see \citealp{RKE2010}) and
estimating the error covariance matrix dynamically using an exponentially
weighted moving average (EWMA). Recently \cite{KK2013b} extended the
methodology to time-varying parameter factor augmented VAR with 
stochastic volatility (TVP-FAVAR-SV). Although
the TVP-VAR-SV are overall easier to handle than TVP-FAVAR-SV in terms of the
online estimation, it remains an open question if a small amount of common
components might efficiently summarize the variation in the data for
forecasting or economic analysis.\\ 
In this paper, we propose a new model
that bridges TVP-VAR-SV and TVP-FAVAR-SV, with a new estimation strategy based
on the results in KK. Specifically, to reduce the dimensionality, we draw from
the recent developments in Multivariate Index Autoregressive (MAI) models, see
\cite{CKM2016}, \cite{CGH2017}, \cite{CG2019}, and \cite{CCM2020}, among
others. The MAI model, originally introduced by \cite{Reinsel1983}, is a
bridge between reduced-rank VARs (see \citealp{CCM2015} and the references
therein) and the Dynamic Factor Model (DFM, see \citealp{SW2016}, 
\citealp{Lippi2019} and the references therein). 
On the one hand, it reduces the dimensionality by
imposing a sort of reduced rank structure to the VAR, on the other, it allows
for identifying few linear combinations of the variables, which are labelled
as the indexes, whose lags are entirely responsible for the dynamics of the
system.\newline Although the mathematical formulation of the MAI is similar to
that of the DFM, an advantage of the former is that it does not require that
the dimension of the system diverges to infinity in estimation. Hence, the MAI
can be applied even to small or medium VARs. Moreover, the factor structure
can be tested for and not simply imposed as in the DFM, and the estimation
error of the indexes is explicitly accounted for, see \cite{CG2019} for
further details.\\
The contribution of the paper is twofold. The first
is to propose a MAI with time-varying parameters and stochastic volatility
(TVP-MAI-SV). A second contribution of the paper is to develop approximate
estimation methods for the TVP-MAI-SV which do not involve the use of Markov
chain Monte Carlo (MCMC) such as in \cite{CCM2018} and \cite{CCM2019}. To
achieve this result we propose mixing the switching algorithm, see
\cite{CGH2017}, with forgetting factors in the same spirit of \cite{KK2013b}.\\
 Forgetting factors (also known as discount factors), have long been
used with state-space models, see \cite{RKE2010}. They do not require the use
of MCMC methods and be useful in economic and financial
applications, see \cite{DH2012}, and \cite{GNS2014}.\\
The new model is applied in two empirical applications. 
The first one is a variance decomposition on a large dataset
composed of 215 time series. 
The analysis of economic uncertainty has a long history. 
A large literature investigates the relationship between uncertainty and growth 
by proving that both at the macro and micro level, uncertainty moves 
counter-cyclically: rising steeply in recessions and falling in booms. Evidence of counter-cyclical volatility is provided, among the others, for macro stock returns in \cite{ScW1989} for firm-level stock asset prices in \cite{Campal01}, for consumption and income in \cite{SCA2004}.
Moreover, given the increase of uncertainty after major economic and political shocks and the recent 2008 financial crisis followed by the Great Recession, the interest of economists and policymakers become markedly focused on its effects on the economy. Moving from the seminal paper of \cite{Bloom2009} that provides a structural framework to analyse the impact of uncertainty shocks, more recent literature starts analysing and 
measuring the macroeconomic and financial uncertainty and its impact on macroeconomic variables (see among the others \citealp{BES2013}, \citealp{CCG2014} and \citealp{Jo2015}).\\
However, measuring uncertainty effect on macroeconomics is difficult as most macro variables move together over the business cycle. This co-movement may face challenging identification problems which are generally overcome by estimating uncertainty in a preliminary step and then evaluating its impact on macroeconomic variables. 
Uncertainty measure is included together with a small set of macroeconomic variables in a VAR model computing the responses of the macro variables to the uncertainty shock (see among others \citealp{Bloom2009}, \citealp{CCG2014}, \citealp{BB2017}, \citealp{BES2013}, \citealp{JSN2015}).
The use of small VAR models, to assess the effects of uncertainty, can make the results subject to the common omitted variable bias and non-fundamentalness of the errors, besides providing results on the impact to just a few economic indicators. Differently from the previous literature, in line with \cite{CKM2016}, we focus on measures of uncertainty based on the volatility and we apply our TVP-MAI-SV on 215 macroeconomic and financial variables.
The log predictive likelihood ($\log$ PL) is used to select among 
the number of indexes that following
\cite{CKM2016}, \cite{GanBreu2010} and \cite{Gugl2016}, are classified 
into 5 groups: \textit{Financial, Labour Market, Nominal, Prices, Real}. 
We analyse how much volatility is explained by each index over time.\\
The second empirical application provides a forecasting exercise for three key macroeconomic variables, namely Real Gross Domestic Product (GDP), Consumer Price Index (CPI) and Effective Federal Funds Rate (IntRate). The point and density forecast evaluation show that the TVP-MAI-SV model has very promising out-of-sample properties compared with a set of univariate and multivariate competitors. In particular, the specification of the model with SV and constant parameters performs significantly well. This is in line with the results in \cite{CE2018} and \cite{CEHK2020}\\
The rest of the paper proceeds as follows. Section
\ref{sec:Model} presents the MAI model and introduces the new TVP-MAI-SV.
Section \ref{sec:BayesEstimation} discusses the new estimation approach.
Section \ref{sec:EmpApplication} contains the empirical application. Finally
Section \ref{sec:Concl} draws some conclusions. All the derivations are reported in Appendix A and B.

\section{From the MAI model to the TV-MAI \label{sec:Model}}

Let $\mathbf{y}_{t}\equiv(y_{1,t},\ldots,y_{N,t})^{\prime}$\ denote the
$N$-vector of the time series of interest. In the fixed parameter framework,
it is assumed that variables $\by_{t}$ are generated by a stationary VAR of
order $p$ (VAR$(p)$):
\begin{equation}
\mathbf{y}_{t}=\Phi(L)\mathbf{y}_{t-1}+\mbox{\boldmath $\varepsilon$}_{t}%
,\hspace{0.3cm}t=1,2,\ldots,T,\label{VAR}%
\end{equation}
where $\Phi(L)=\sum_{h=0}^{p-1}\Phi_{h}L^{h}$ and
$\mbox{\boldmath $\varepsilon$}_{t}$ are i.i.d. innovations with
$\mbox{E}(\mbox{\boldmath $\varepsilon$}_{t})=0$,
$\mbox{E}(\mbox{\boldmath $\varepsilon$}_{t}\mbox{\boldmath $\varepsilon$}_{t}%
^{\prime})=\mbox{H}$ (positive definite) and finite fourth moments.\newline In
order to reduce the number of parameters of model (\ref{VAR}),
\cite{Reinsel1983} proposed to impose the following set of restrictions on the
VAR mean parameters:
\begin{equation}
\Phi(L)=\beta(L)\omega^{\prime},\label{assumption}%
\end{equation}
where $\omega$ is full-rank $N\times q\ $matrix with$\ q<N,$\ $\beta
(L)=\sum_{h=0}^{p-1}\beta_{h}L^{h}$,\ and $\beta_{h}$\ is a $N\times
q\ $matrix\ for $h=1,\ldots,p$.\newline The rationale underlying assumption
(\ref{assumption}) is that the unrestricted VAR foresees $N$ linearly
independent mechanisms by which past information is transmitted to the system.
However, since it is commonly believed that few common shocks generate most of
macroeconomic fluctuations, it is reasonable to assume that there is a reduced
number of channels through which variables are influenced by their past. In
other words, this is exactly what Equation (\ref{assumption}) states (see
\citealp{CKM2016} and \citealp{CG2019} for further details).\newline Notice that
Assumption in Equation (\ref{assumption}) is equivalent to postulating the
following structure for series $\mathbf{y}_{t}$:
\begin{equation}
\mathbf{y}_{t}=\beta(L)\mathbf{f}_{t-1}+\mbox{\boldmath $\varepsilon$}_{t}%
,\label{MAI}%
\end{equation}
where $\mathbf{f}_{t}=\omega^{\prime}\mathbf{y}_{t}$. \cite{Reinsel1983}
defines the $q$-dimensional series $\mathbf{f}_{t}=(f_{1,t},\ldots,f_{q,t})$
as the index variables and labels equation (\ref{MAI}) as the MAI
model.\newline An interesting property of the MAI is that the indexes
themselves have a VAR$(p)$ representation. Indeed, if we premultiply by
$\omega^{\prime}$ both sides of equation (\ref{MAI}) we get%
\[
\mathbf{f}_{t}=\alpha(L)\mathbf{f}_{t-1}+\mbox{\boldmath $\epsilon$}_{t},
\]
where $\alpha(L)=\omega^{\prime}\beta(L)$ and $\mbox{\boldmath $\epsilon$}_{t}%
=\omega^{\prime}\mbox{\boldmath $\varepsilon$}_{t}$. This feature is in sharp
contrast with reduced rank VAR models, where linear combinations of the
variables do not generally admit a finite order VAR representation, see
\cite{CHP2009}, and highlights the analogy between the role of the indexes in
the MAI and the factors in DFMs.\newline Recently, there has been a
renewed interest in the MAI, \cite{CKM2016} derived classical and Bayesian
estimation of large MAIs and applied this modelling for structural analysis,
\cite{CGH2017} proposed a multivariate realized volatility model that is
endowed with an index structure. \cite{CG2019} extended the model by allowing
for individual AR structures, and \cite{CCM2020} offered a MAI with stochastic
volatility and provide MCMC estimation.\newline We extend the traditional MAI
model allowing the variation in both the mean and variance equation, the
TVP-MAI-SV takes the form:
\begin{equation}
\mathbf{y}_{t}={\textstyle\sum\limits_{j=1}^{q}}%
\mbox{\boldmath $\beta$}(L)_{j,t}f_{j,t-1}+\mbox{\boldmath $\varepsilon$}_{t}%
,\hspace{0.5cm}\mbox{\boldmath $\varepsilon$}_{t}\sim \N(0,\mbox{H}_{t}%
),\label{eq:TVIAAR}%
\end{equation}
where $f_{j,t}=\sum_{k=1}^{N}\omega_{k,j}y_{k,t}$, and
$\mbox{\boldmath $\beta$}(L)_{j,t}$ is a polynomial $N-$vector of time varying
coefficients that evolve as random walks for $j=1,\ldots,q$. \ Notice that,
similarly in the literature on the TVP-FAVAR, we assume that the loadings
of the indexes vary over time whereas the weights $\omega$\ remain stable.
Finally the $\mbox{\boldmath $\varepsilon$}_{t}$ follows a multivariate
stochastic volatility model given by $\mbox{H}_{t}$.\newline The model given
in Equation \eqref{eq:TVIAAR} is difficult to estimate with already existing
methods, and to tackle this task we develop a new hybrid algorithm that is
described below.

\section{Estimation \label{sec:BayesEstimation}}

The estimation of model in Equation \eqref{eq:TVIAAR} is based in a fast
two-step algorithm which vastly reduces the computational burden. Subsection
\ref{subsec:Bayes} presents the state space representation of TVP-MAI-SV and
briefly explains the related estimation issues. Subsection \ref{subsec:Hybrid}
presents the new hybrid switching algorithm used to estimate the TVP-MAI-SV.
Subsection \ref{subsec:Predictive} describes model selection. All the
derivations are reported in Appendix A.

\subsection{Bayesian Estimation \label{subsec:Bayes}}
The model in Equation \eqref{eq:TVIAAR} can be casted in state space form as follows:
\begin{equation}%
\begin{split}
\mathbf{y}_{t} &  =\mathbf{Z}_{t}\mbox{\boldmath $\beta$}_{t}%
+\mbox{\boldmath $\varepsilon$}_{t},\hspace{0.5cm}%
\mbox{\boldmath $\varepsilon$}_{t}\sim \N(0,\mathbf{H}_{t}),\\
\mbox{\boldmath $\beta$}_{t} &  =\mbox{\boldmath $\beta$}_{t-1}%
+\mbox{\boldmath $\eta$}_{t},\hspace{0.5cm}\mbox{\boldmath $\eta$}_{t}%
\sim \N(0,\mbox{Q}),
\end{split}
\label{eq:ModelloSSTV}%
\end{equation}
where $\mathbf{y}_{t}$ is the vector of observed time series at time $t$, 
$\mathbf{Z}_{t}=I_N \otimes (f_{1,t},\ldots, f_{q,t})$ is the stack of
all the indexes depending on the unknown $\mbox{\boldmath $\omega$}$,
$\mbox{\boldmath $\beta$}_{t}=(\beta'_{1,t},\ldots,\beta'_{q,t})'$
is an $Nq\times1$ vector containing the time varying $\mbox{\boldmath $\beta$}$s
(states), which are assumed to follow a random walk dynamic. Finally the
errors, $\mbox{\boldmath $\varepsilon$}_{t}$ and $\mbox{\boldmath $\eta$}_{t}$ 
are assumed to be
mutually independent at all leads and lags and $\mathbf{H}_{t}$ features
stochastic volatility.\\
The model in equation
\eqref{eq:ModelloSSTV} is used in a number of recent paper, see among others
\cite{Primiceri2005}, \cite{KoopLeonStrachan2009} and \cite{KK2012}.
Traditionally, it is estimated with both classical and Bayesian approaches. In
the first case, the likelihood is efficiently calculated with the Kalman
filter (KF) routine, see \cite{DurbinKoopman2001}, and the time-varying
parameters are automatically filtered as latent state variables, once that
$\mbox{H}_{t}$ and $\mbox{Q}$ are estimated.\newline The Bayesian estimation
method, which requires simulation based methods, such as the MCMC, involves
the specification of $\mbox{H}_{t}$ and $\mbox{Q}$ together with the initial
condition, $\mbox{\boldmath $\beta$}_{0\mid0}$, of the model parameters,
for an introduction see \cite{Koop2003}. Although Bayesian algorithms are reliable
in this context, as recently discussed in \cite{CCM2018}, they become
computational intensive as the number of parameters increases and they become
unfeasible when a large amount of models (e.g. different number of factors)
have to be estimated.\newline To solve this issue we propose a new hybrid
estimation technique, that uses the discount factor methodology proposed by
\cite{RKE2010} and \cite{KK2012} (see Appendix A), to estimate
$\mbox{\boldmath $\beta$}_{t}$ and $\mbox{H}_{t}$, and a switching algorithm
to estimate $\mbox{\boldmath $\omega$}$.

\subsection{Hybrid algorithm for TVP-MAI-SV \label{subsec:Hybrid}}

The model in equation \eqref{eq:ModelloSSTV}, has both static
($\mbox{\boldmath $\omega$}$) and dynamic parameters
($\mbox{\boldmath $\beta$}_{t}$). Following \cite{CGH2017} and \cite{KK2013b}
we combines the ideas of variance discounting methods with the switching
algorithm in order to obtain analytical results for the posteriors of the
``indexes" ($\mbox{\boldmath $\beta$}_{t}$) as well as the static parameters
($\mbox{\boldmath $\omega$}$). The model also features stochastic volatility
and the algorithm needs to take it into account in the $\mathbf{H}_{t}$. The
$\mathbf{H}_{t}$ can be easily estimated using the EWMA filter.\newline The
estimation starts from the same algorithm described in \cite{CGH2017} and
introduces the time-varying parameters as follows:

\begin{enumerate}
\item[1)] Given (initial) estimates of $\mbox{\boldmath $\omega$}$ and
$\mathbf{H}_{t}$, run the KF for model reported in equation \eqref{eq:ModelloSSTV}
to get the optimal estimates of the latent states $\hat
{\mbox{\boldmath $\beta$}}_{t}=(\hat{\beta}_{1,t},\ldots,\hat{\beta}_{Nq,t})$,
see Appendix A;
\item[2)] Given the $\hat{\mbox{\boldmath $\beta$}}_{t}$ and the
$\mbox{\boldmath $\omega$}$, extract $\mathbf{H}_{t}$ using an EWMA estimator
for the measurement error covariance matrix:
\[
\hat{\mathbf{H}}_{t} = \kappa\hat{\mathbf{H}}_{t-1} + (1-\kappa)\hat
{\mbox{\boldmath $\varepsilon$}}_{t}\hat{\mbox{\boldmath $\varepsilon$}}%
_{t}^{^{\prime}},
\]
where $\hat{\mbox{\boldmath $\varepsilon$}}_{t} = \mathbf{y}_{t} -
\mbox{\boldmath $\beta$}_{t}\mathbf{Z}_{t}$ is given as output from the Kalman
filter. The EWMA decay factor $\kappa$ requires to be selected, we discuss
this issue in Section \ref{sec:EmpApplication}.
\item[3)] Premultiply by $\hat{\mathbf{H}}_{t}^{-1/2}$ and apply the
$\mathrm{Vec}$ operator to both the sides of Equation (\ref{eq:TVIAAR}), then
use the property $\mathrm{Vec}(\mbox{A} \mbox{B} \mbox{C})=(\mbox{C}^{\prime
}\otimes\mbox{A})\mathrm{Vec}(\mbox{B})$ to get:
\begin{equation}
\mathrm{Vec}(\hat{\mathbf{H}}_{t}^{-1/2}\mathbf{y}_{t})={\textstyle\sum
\limits_{h=1}^{p-1}}(\mathbf{y}_{t-h}^{\prime}\otimes\hat{\mathbf{H}}_{t}%
^{-1/2}\hat{\beta}_{h,t})\mathrm{Vec}(\mbox{\boldmath $\omega$}^{\prime
})+\mathrm{Vec}(\hat{\mathbf{H}} _{t}^{-1/2}\mbox{\boldmath $\varepsilon$}_{t}%
). \label{eq:Vec-model}%
\end{equation}
Given the previously obtained estimates of $\mbox{\boldmath $\hat{\beta}$}_{t}%
$ and $\hat{\mathbf{H}}_{t}$, estimate $\mathrm{Vec}%
(\mbox{\boldmath $\omega$}^{\prime})$\ with OLS in equation
(\ref{eq:Vec-model}).
\item[4)] Repeat steps 1, 2 and 3 till numerical convergence occurs.
\end{enumerate}
Few comments are in order. The above algorithm offers several
advantages over the available alternatives, including computational
simplicity, no need of a normalization condition on the parameters
$\mbox{\boldmath $\omega$}$, over-identifying restrictions can be easily
imposed on $\mbox{\boldmath $\omega$}$ and optimization is explicit at each
step. In order to speed up numerical convergence, it is important to take
proper choices regarding various hyperparameters and initial conditions. As in
\cite{KK2013b}, we choose fairly non-informative priors. The initial
conditions for the time-varying parameters $\mbox{\boldmath $\beta$}_{t}$ and
the time-varying covariance $\mathbf{H}_{t}$ are set as follows:
$\mbox{\boldmath $\beta$}_{0}\sim\mbox{N}(0,4)$ and $\mathbf{H}_{0}%
=\mbox{I}_{n}$.
Finally the initial conditions for $\mbox{\boldmath $\omega$}$ are obtained
from the eigenvectors that are associated with the first $q$ principal
components of series $\mathbf{y}_{t}$.

\subsection{Dynamic model averaging and Dynamic model selection for the TVP-MAI-SV \label{subsec:Predictive}}

The model discussed in Section \ref{subsec:Bayes} is just one of the
$\mathcal{M}$ possible models. Depending on the selection of several parameters (number of
factors, forgetting factor, decay factor $\lambda$, time varying parameters
$\bbeta$), many models can be switched over time. Considering the
$\mathcal{M}$ possible models equation \eqref{eq:ModelloSSTV} can be written as follows:
\begin{equation}%
\begin{split}
\mathbf{y}_{t}  &  = \mathbf{Z}_{t}^{(k)}\mbox{\boldmath $\beta$}_{t}^{(k)} +
\mbox{\boldmath $\bvarepsilon$}_{t}^{(k)}, \hspace{0.5cm }%
\mbox{\boldmath $\bvarepsilon$}_{t}^{(k)}\sim \N(0, \bH_{t}^{(k)}),\\
\mbox{\boldmath $\beta$}_{t}^{(k)}  &  = \mbox{\boldmath $\beta$}_{t-1}^{(k)}
+ \mbox{\boldmath $\eta$}_{t}^{(k)}, \hspace{1.1cm}\mbox{\boldmath $\eta$}_{t}^{(k)}%
\sim \N(0, \mbox{Q}^{(k)}),
\end{split}
\label{eq:PDModelloSSTV}%
\end{equation}
where $\mathbf{Z}_{t}^{(k)}$ and $\mbox{\boldmath $\beta$}_{t}^{(k)}$ are, respectively, one of the possible set of indexes and
time-varying parameters.\newline Looking at equation \eqref{eq:PDModelloSSTV} is clear that
there are potentially a large number of models at each time point $t$. When
faced with multiple models, it is common to use model selection or model
averaging techniques, that in our framework have to be dynamic. More
specifically in a model selection exercise, we want to allow for the selected
model to change over time, thus doing Dynamic Model Selection (DMS). 
In a model averaging exercise, we
want to allow for the weights used in the averaging process to change over
time, thus leading to Dynamic Model Averaging (DMA). 
In this paper, we do both using the same approach
of \cite{RKE2010}, see Appendix A.\\ 
In DMS and DMA the main objective
is to calculate $\pi_{t|t-1,j}$ which is the probability that model $j$
applies at time $t$, given information through time $t-1$. Once $\pi
_{t|t-1,j}$ for $j= 1, \dots, J$ are obtained they can either be used to do
model averaging or model selection. \newline DMS arises if, at each point in
time, the model with the highest value for $\pi_{t|t-1,j}$ is used. Note that
$\pi_{t|t-1,j}$ will vary over time and, hence, the selected model may switch
over time. DMA arises if model averaging is done in period $t$ using
$\pi_{t|t-1,j}$ for $j=1, \dots, J$ as weights. The contribution of
\cite{RKE2010} is to develop a fast recursive algorithm for calculating
$\pi_{t|t-1,j}$ that, given an initial condition: $\pi_{0|0,j}$ for $j=1,
\ldots, J$, derives a model prediction equation using the forgetting factor
$\alpha$:
\[
\pi_{t|t-1,j} = \dfrac{\pi_{t-1|t-1,j}^{\alpha}}{\sum_{j=1}^{J}\pi
_{t-1|t-1,j}^{\alpha}},
\]
and a model updating equation of
\[
\pi_{t|t,j} = \dfrac{\pi_{t|t-1,j}f_{j}(\mbox{y}_{t}|\mbox{y}_{1:t-1})}%
{\sum_{j=1}^{J}\pi_{t|t-1,j}f_{j}(\mbox{y}_{t}|\mbox{y}_{1:t-1})},
\]
where $f_{j}(\mbox{y}_{t}|\mbox{y}_{1:t-1})$ is the predictive likelihood of
model $j$. The $0<\alpha\leq1$ is a forgetting factor that tunes the frequency
of switch between models occurred over time. Low values of $\alpha$
corresponds to a rapid switch, high values give the opposite. Naturally $\alpha=
1$ brings to the conventional Bayesian Model Averaging (BMA).\newline Finally,
the initial condition is set to the equal probability $\pi_{0|0,j}=1/J,
\hspace{0.1cm} \forall j$.

\section{Empirical application \label{sec:EmpApplication}}

We use the new TVP-MAI-SV to carry out both a structural and a forecasting
exercise.\newline Subsection \ref{subsec:DataDescription} presents the dataset
considered in our study. Structural Analysis is described in subsection
\ref{subsec:VarDec}. While, subsection \ref{subsec:Forecast} discusses the
forecasting exercise.

\newpage
\subsection{Data description \label{subsec:DataDescription}}
The quarterly data used in the paper are download from Fred-Database and they run from 1960:1 to 2019:4. Following \cite{CKM2016}, \cite{GanBreu2010} and \cite{Gugl2016} and references therein, we classify the series into 5 groups: Real (RI), 
Nominal (NI), Labour Market (LMI), Prices (PI) and Financial (FI) indexes. 
A detailed description of all the 215 series as listed in each group are
reported from Table \ref{t:fc1} to Table \ref{t:fc6} in Appendix B. All the
variables are transformed to achieve stationarity and then
standardized as specified in \cite{MCNg2020}.
\newline For the structural exercise we use all the 215 series,
for the forecasting exercise we follow KK and we use a selection of 25 series. Those are
highlighted in bold in the Tables \ref{t:fc1} to \ref{t:fc6} of Appendix B.

\subsection{Variance Decomposition \label{subsec:VarDec}}

Following \cite{CCM2020} we carry out a variance
decomposition analysis using the full dataset
described in Appendix B. Before using the TVP-MAI-SV we proceed
to select some of its key features such as: the number of indexes ranging from 1 to 5; different values for $\lambda = \{0.96, 0.97, \ldots, 1\}$; and $\kappa = \{0.94, 0.96, \ldots, 1\}$. This provides a total of 80 alternative specifications. 
Regarding the lag length we select 4 lags, see \cite{KK2013} and \cite{KMV2019b}. We estimate all these specifications
and rank them according to the $\log$ PL, as discussed in \cite{BCRvD2016} and
 computed as discussed in Appendix A.\\
Table \ref{tab:modelsel} provides results for the best 5 specifications together with the worst 5 specifications we found over the total 80 specifications we searched over. The table
contains the number of factors and the values of $\lambda$ and $\kappa$ that
uniquely identifies a specification. Looking at the table the best specification 
contains 5 indexes, $\lambda = 0.99$ and a $\kappa = 0.94$. This parameter combination 
features a $\log$ PL equal to 488.1631.
\begin{table}[h]
\caption{The Table reports: the model 
ranking (Ranking); the number of indexes (\#Factors); the shrinkage parameter for the
time-varying parameters ($\lambda$); the decay factor for the EWMA estimator ($\kappa$); and 
the log-predictive likelihood for each model ($\log$ PL). The
second, third and forth columns contain the factor,$\lambda$-$\kappa$ combination that
uniquely identifies a specification.}
\label{tab:modelsel}
\centering{
\begin{tabular}[c]{c|c|c|c|r}%
\toprule
Ranking  & \# Factors & $\lambda$ & $\kappa$ & $\log$ PL\\
\midrule
1   & 5 & 0.99 & 0.94 & 488.1631  \\
2   & 4 & 1.00 & 0.94 & 487.7487  \\
3   & 2 & 0.97 & 0.94 & 481.1135  \\
4   & 3 & 1.00 & 0.94 & 478.4866  \\
5   & 2 & 0.99 & 0.94 & 478.8624  \\
\dots & \dots & \dots & \dots & \dots \\
76  & 4 & 0.97 & 1.00 & -242.8570 \\
77  & 2 & 0.97 & 1.00 & -243.4300 \\
78  & 3 & 0.96 & 1.00 & -246.4240 \\
79  & 4 & 0.96 & 1.00 & -250.2354 \\
80  & 5 & 0.96 & 1.00 & -258.7879 \\
\midrule
\end{tabular}}
\end{table}
\begin{figure}[h]
\caption{Estimated Indexes. NBER recession are reported with grey
vertical bar. The name of the factor is also reported.}%
\label{fig:Factors}
\vspace{-0.2cm} 
\hspace{-0.5cm}
\includegraphics[width=16.5cm,height=10cm]{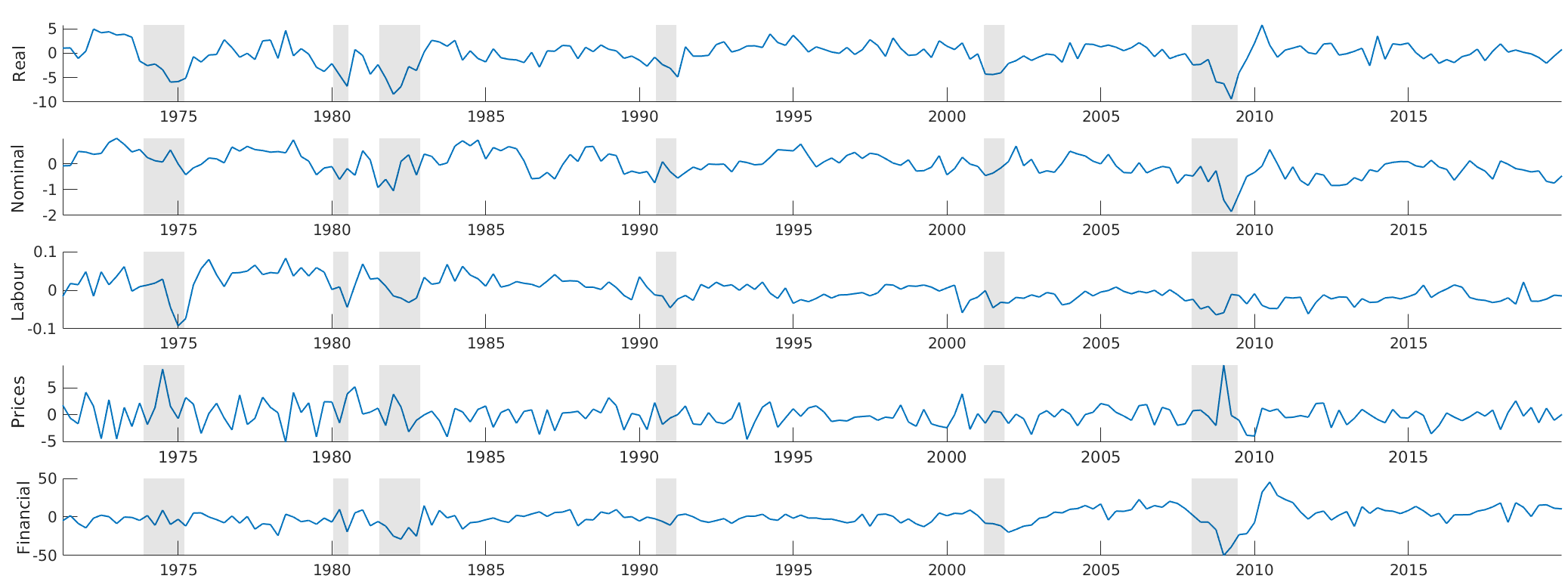}
\end{figure}
Figure \ref{fig:Factors} reports the five estimated indexes.\\ 
Generally speaking, indexes capture accurately both the Oil crisis in 1970s and the Great Financial Crisis of 2008. In more detail, it is evident the effect of the first oil Crisis in 1973 in which prices increased 400\%, is well captured by the pick of the related Price Index (PI) at the beginning of the recession period, followed by a decreasing effect in Real Index (RI), Nominal Index (NI) and above all Labour Market Index (LMI).
Differently, at the beginning of 1979, the second shock evolved more slowly, as producers, led by the Organization of Petroleum Exporting Countries (OPEC), affirmed the concept of setting oil prices and establishing production quotas.
Consequently, if we see an increase in the prices factor, the decreasing effect on the other components arrives later on in 80s. It is interesting that the financial crisis is well displayed by all the factors. Indeed, we appreciate an increase in the financial market associated with a large decrease of both real, nominal factors and an increase in prices. While we notice a slower decrease in the labour trend whose fully recovery seems to come long after.\\ 
The RI has a large volatility spike after the 1970s, observed in correspondence with the last Financial Crisis. The NI shows a smaller degree of time variation than the RI. The FI shows that financial
uncertainty spikes during the recession with the Great Financial Crisis that
dominates the others.
\begin{figure}[!h]
\vspace{-3cm} \centering
\subfloat[Real Gross Domestical Product]{\includegraphics[width=0.48\textwidth]{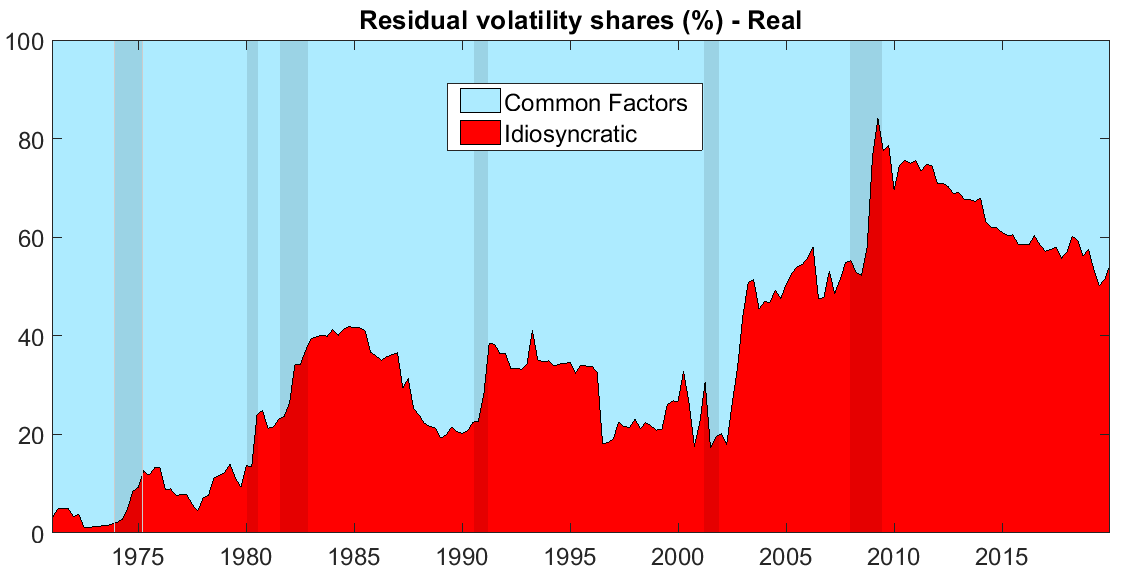}}\quad
\subfloat[TB-3M]{\includegraphics[width=0.48\textwidth]{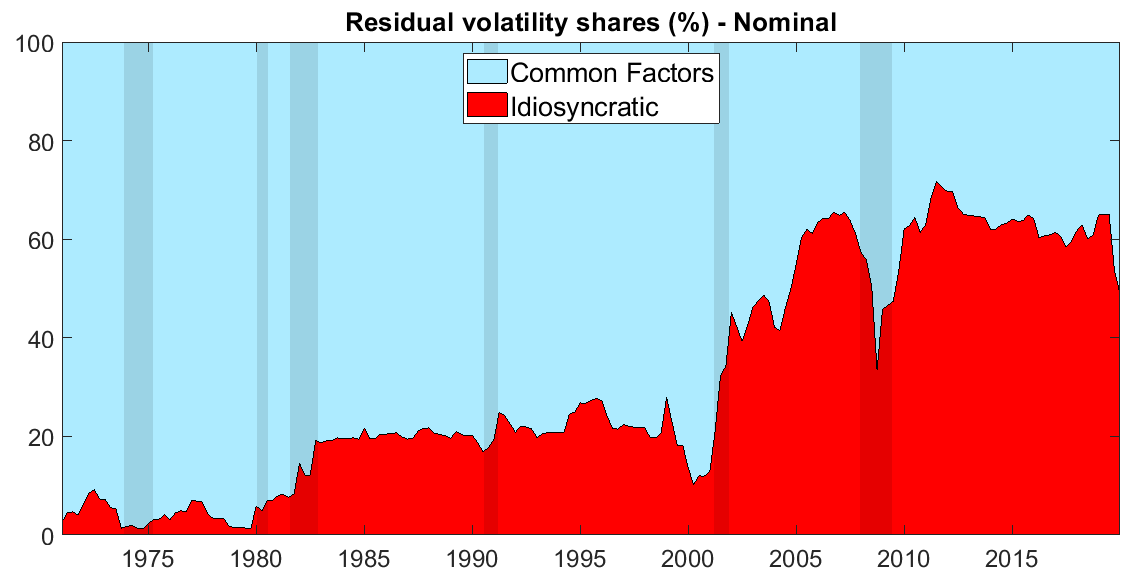}}\newline%
\subfloat[Nonfarm Payrolex]{\includegraphics[width=0.48\textwidth]{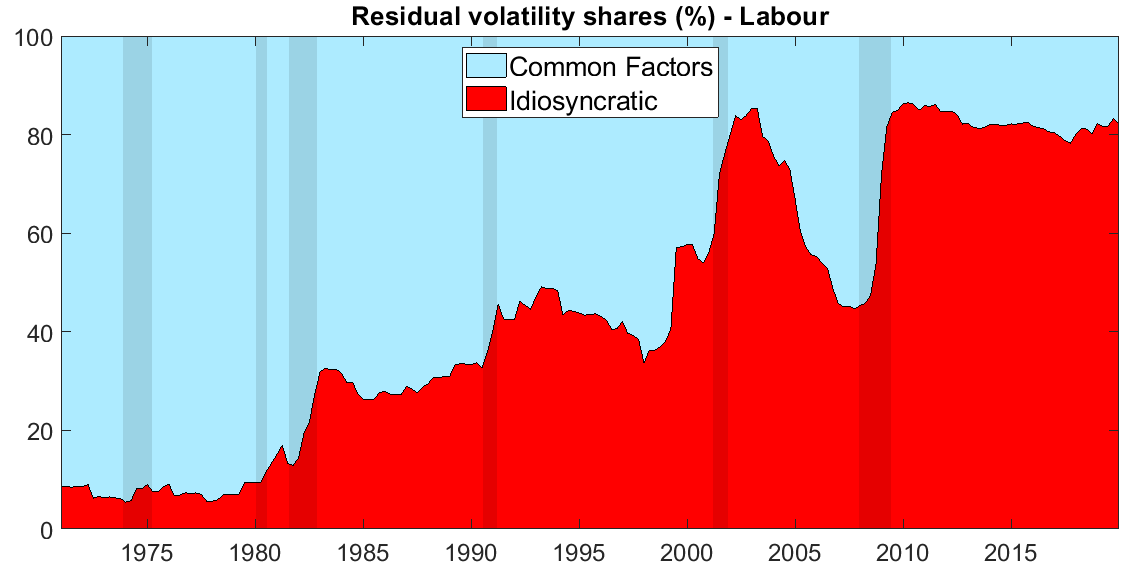}}\quad
\subfloat[CPI Inflation]{\includegraphics[width=0.48\textwidth]{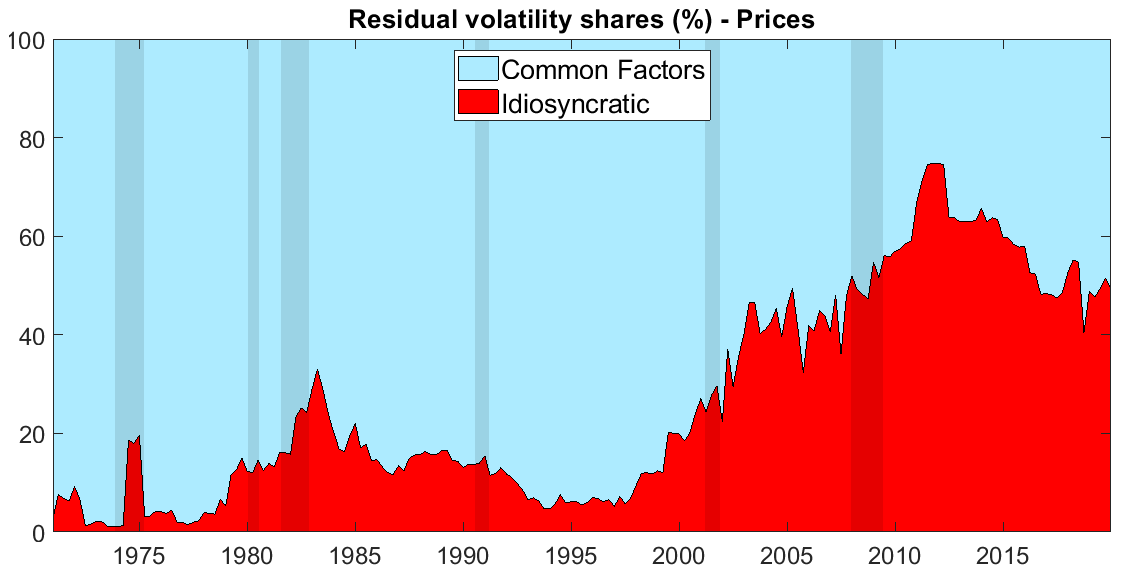}}\newline%
\subfloat[FedFunds]{\includegraphics[width=0.48\textwidth]{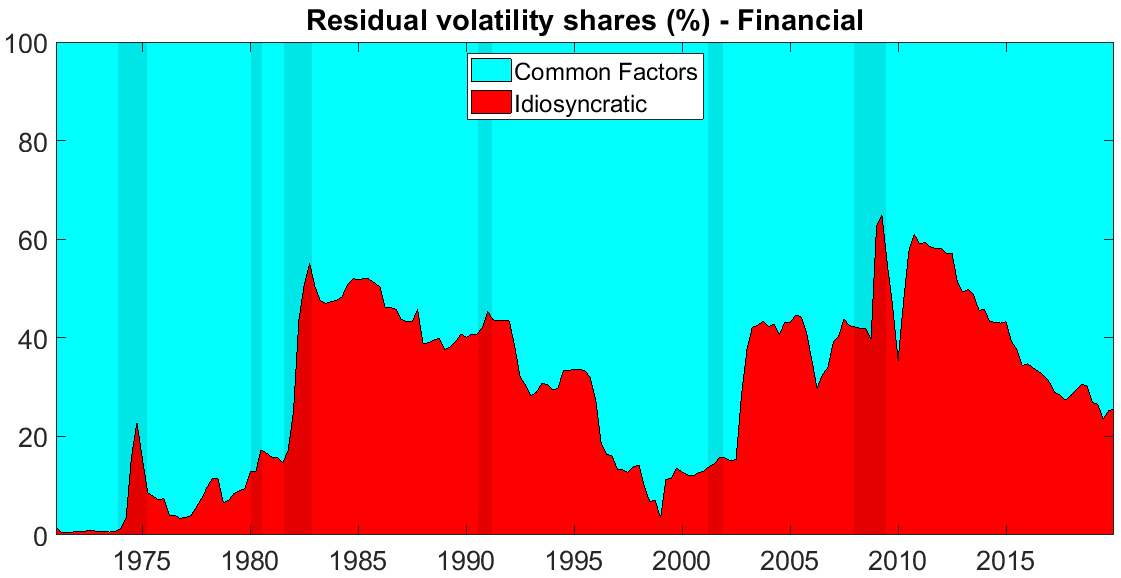}}\quad
\subfloat[Fanchart of all the 217 serie.]{\includegraphics[width=0.48\textwidth]{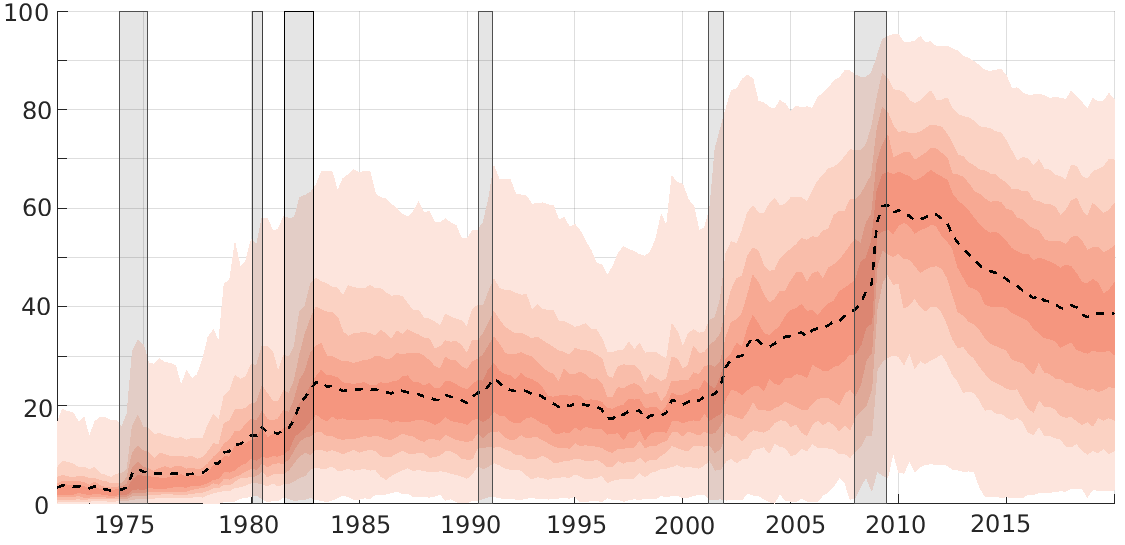}}\newline
\caption{Volatility shares (\%): Common (red), Idiosyncratic (blue).
The NBER recessions are reported with grey vertical lines. Each plot correspond
to the first series reported in Tables \ref{t:fc1} to Table \ref{t:fc6} in
Appendix B. (a) Real Gross Domestic Product, first series in Table \ref{t:fc1}
Appendix B. (b) 3-Month Treasury Bill first series in Table \ref{t:fc3}
Appendix B. (c) Nonfarm payroll first series in Table \ref{t:fc4} Appendix B.
(d) Consumer price index first series in Table \ref{t:fc5} Appendix B. (e)
Federal Funds rate first series in Table \ref{t:fc6} Appendix B. (f) Fanchart
of all the 217 series, with the mean as black(dotted) line.}%
\label{fig:VarianceDecomposition}%
\end{figure}
To analyse the explained volatility of each component we
report in Figure \ref{fig:VarianceDecomposition} the time-varying percentage
shares of explain volatility by the idiosyncratic and common components. The
Figure reports just the first series of Tables \ref{t:fc1} to \ref{t:fc5} in
Appendix B.\newline Starting from the RI and the corresponding series GDP, it
seems that the common component of volatility has increased its importance
during 2000 with a big increase after the Great Recession. The fraction of
volatility explained by the common component is much higher for the Nonfarm
payroll, first series of LMI.\\
Moving to the FI and for the first series
that is the Federal funds rate, the common component is quite high during the
considered period. Interesting the percentage of common component change a lot
with big increase in correspondence of the economic crisis.

\subsection{Forecasting exercise \label{subsec:Forecast}}

This section provides the out-of-sample performance of the TVP-MAI-SV against a set of standard
competitors.\newline In this exercise we consider the 25 major quarterly U.S.
macroeconomic variables as discussed in KK. The series are reported in bold
from Table \ref{t:fc1} to Table \ref{t:fc5} in Appendix B. We focus on
empirical results relating to three variables: CPI inflation, GDP growth and
the Federal Funds rate and refer to these as the main variables.\newline The
forecasting exercise is performed using an expanding window with an initial
estimation sample runs from 1960:Q3 to 1972:Q4. The model is then recursively
estimated on a forecast windows of 183 quarterly vintages (forecasting windows
start from 1973:Q1 through 2020:Q1).\\
Imposing simple restriction, the
TVP-MAI-SV of equation \eqref{eq:ModelloSSTV}, encompasses some multivariate models:
\begin{itemize}
\item[1)] The original MAI model, as in \cite{Reinsel1983} and \cite{CKM2016}
when $\mbox{\boldmath $\beta$}_{t} = \mbox{\boldmath $\beta$}_{t-1}$ and
$\bH_{t}$ are time invariant ($\mbox{Q} = 0$, $\kappa= 1$ and
$\bH_{t}$ set to the OLS estimates $\forall t$).

\item[2)] The MAI-SV similar to \cite{CCM2018} when
$\mbox{\boldmath $\beta$}_{t} = \mbox{\boldmath $\beta$}_{t-1}$ is time
invariant ($\mbox{Q}=0$) but $\kappa= 1$ and $\bH_{t}$ evolves over time.

\item[3)] The TVP-MAI model without stochastic volatility when
$\mbox{\boldmath $\beta$}_{t}$ is time varying ($\mbox{Q}\neq0$) but
$\bH_{t}$ is fixed and set to the OLS estimates.
\end{itemize}
In addition to the models discussed previously, Table
\ref{tab:Multivariatemodels} reports all the competitive models considered in
our forecasting analyses.\newline\begin{table}[h]
\caption{The table reports all the models considered in the forecasting
exercise plus the benchmark model. The first column is the abbreviation of the
model. The second column provides a brief description of each model.}%
\label{tab:Multivariatemodels}
{\tiny \setlength\tabcolsep{3.0pt}{}
\begin{tabular}
[c]{l|l}%
\toprule \textit{Abbreviation} & \textit{Full Description}\\
\midrule $\mathcal{M}_{1}$ & TVP-MAI-SV. Number of indexes and the optimal
value of the $\lambda$ and $\kappa$\\
& are selected using DMA as outlined in \cite{KK2013}. In this model $\alpha=
0.99$.\\
\midrule $\mathcal{M}_{2}$ & TVP-MAI-SV. Number of indexes and the optimal
value of the $\lambda$ and $\kappa$\\
& are selected using DMS as outlined in \cite{KK2013}. In this model $\alpha=
0.99$.\\
\midrule $\mathcal{M}_{3}$ & MAI-SV, with fix $\beta_{t}$($\lambda= 1$).
Number of indexes and the optimal value of $\kappa$\\
& are selected using DMA as outlined in \cite{KK2013}. In this model $\alpha=
0.99$.\\
\midrule $\mathcal{M}_{4}$ & MAI-SV, with fix $\beta_{t}$($\lambda= 1$).
Number of indexes and the optimal value of $\kappa$\\
& are selected using DMS as outlined in \cite{KK2013}. In this model $\alpha=
0.99$.\\
\midrule $\mathcal{M}_{5}$ & TVP-MAI, with fix $\mbox{H}$ ($\kappa= 1$).
Number of indexes and the optimal value of the $\lambda$\\
& are selected using DMA as outlined in \cite{KK2013}. In this model $\alpha=
0.99$.\\
\midrule $\mathcal{M}_{6}$ & TVP-MAI, with fix $\mbox{H}$ ($\kappa= 1$).
Number of indexes and the optimal value of the $\lambda$\\
& are selected using DMS as outlined in \cite{KK2013}. In this model $\alpha=
0.99$.\\
\midrule $\mathcal{M}_{7}$ & MAI. Number of indexes are selected using DMA as
outlined in \cite{KK2013}. In this model $\alpha= 0.99$.\\
\midrule $\mathcal{M}_{8}$ & MAI. Number of indexes are selected using DMS as
outlined in \cite{KK2013}. In this model $\alpha= 0.99$.\\
\midrule $\mathcal{M}_{9}$ & Random Walk.\\
\midrule $\mathcal{M}_{10}$ & Vector Autoregressive(1) estimated using the
OLS.\\
\midrule $\mathcal{M}_{11}$ & Vector Autoregressive(4) estimated using the
OLS.\\
\midrule $\mathcal{M}_{12}$ & Dynamic Factor Model\\
\midrule $\mathcal{M}_{13}$ & TVP-VAR-SV with 4 lags and stochastic
volatility. Optimal value of the shrinkage parameter is selected\\
& using DMS as outlined in \cite{KK2013}. In this model $\lambda= 0.99$,
$\kappa= 0.98$ and $\alpha= 0.99$. \textbf{Benchmark model}\\
\midrule $\mathcal{M}_{14}$ & TVP-VAR-SV with 4 lags and stochastic
volatility. Optimal value of the shrinkage parameter is selected\\
& using DMS as outlined in \cite{KK2013}. In this model $\lambda$ is
dynamically selected\\
& and $\kappa= 0.98$ and $\alpha= 0.99$.\\
\midrule $\mathcal{M}_{15}$ & TVP-FAVAR-SV with 4 lags and 4 indexes as in
\cite{KK2013b}.\\
\bottomrule
\end{tabular}
}\end{table}The benchmark model is a TVP-VAR with four lags ($\mathcal{M}%
_{15}$), following KK, the optimal Minnesota shrinkage coefficient ($\gamma$)
is set to 0.005. We also include VAR with 1 and 4 lags estimated using OLS, a
DFM and a random walk process(RW).\\
The TVP-MAI-SV require to specify
the number of indexes, and the values of $\lambda$, $\kappa$ and $\alpha$. We
consider the $q = \{1, 2, 3\}$ indexes and a range of values for the forgetting
factor, $\lambda\in\{0.97, 0.98, 0.99, 1\}$ covering from rapid coefficient
change to no change. For the decay factor, we consider the grid of values
$\kappa\in\{0.96, 0.98, 1\}$. The number of indexes and the values of
$\lambda$, and $\kappa$ are selected dynamically using DMS or DMA, see Table
\ref{tab:Multivariatemodels}. We fix the $\alpha$ to 0.99, other values are
possible and the results are available from the authors upon request.\newline We
assess the performance of our forecasts accuracy in term of point and density
forecast following the evaluation framework of \cite{CEHK2020}. Let $y_{t}%
^{*}$ denote the random variables being forecast and $\hat{y}_{t}$ be their
realizations and the root mean squared forecast error (RMSFE) and mean
absolute forecast error (MAFE) given by:
\begin{equation}%
\begin{split}
\mbox{RMSFE}_{j, h}^{k}  &  = \sqrt{\frac{\sum_{t=t_{0}}^{T-h}y_{t+h}^{*}
-\hat{y}_{j, t+h}^{k}}{T - h - t_{0} + 1}},\\
\mbox{MAFE}_{j, h}^{k}  &  = \frac{\sum_{t=t_{0}}^{T-h} \left|  y_{t+h}^{*} -
\hat{y}_{j, t+h}^{k}\right|  }{T - h - t_{0} + 1}.
\end{split}
\end{equation}
where $k = \{\mathcal{M}_{1}, \ldots, \mathcal{M}_{15}\}$ is the model set, $h =
\{1, \ldots, H\}$ are the forecasting step ahead and $j = \{1, \ldots, 3\}$
are the main variables.\newline To evaluate the density forecasts, we use
the average log-predictive likelihood (ALPL) as described in \cite{Korobilis2019}
and \cite{CEHK2020} as a broadest measure of density accuracy, see also
\cite{Geweke2005}:
\begin{equation}
\mbox{ALPL}_{j, h}^{k} = \frac{\log p_{t+h} (y_{t+h}^{*} - \hat{y}_{j,
t+h}^{k}) }{T - h - t_{0} + 1}.\\
\end{equation}
Table \ref{tab:GDP} to Table \ref{tab:FedFunds} report the ratios of each
model's RMSFE and MAFE with respect to the benchmark model. Entries smaller
(bigger) than 1 indicate that the given model yields forecasts that are more
(less) accurate than those from the baseline. The Tables also report the ALPL
relative to the benchmark model. Values of ALPL higher (lower) than 1 signify
better (worse) performance than the benchmark.

\begin{sidewaystable}
	\caption{Point and density forecast results for GDP. Root Mean Squared Forecast Error (RMSFE) upper panel, Median Absolute Forecast Error (MAFE),
		middle panel, Average Log Predictive Likelihood (ALPL) bottom panel. Results are reported relative to the benchmark specification
		($M_{13}$) for which the values is equal to 1, RMSFE-MAFE lower (higher) than 1 signify
		better (worse) performance than the benchmark. ALPL higher (lower) then 1 signify better (worse) performance. The description of the model is
		reported in Table \ref{tab:Multivariatemodels}. $---$ indicates divergence of the forecast.}%
	\label{tab:GDP}%
	\centering
	{\scriptsize {
			\begin{tabular}
				[c]{l|c|c|c|c|c|c|c|c|c|c|c|c|c|c|c}%
				\toprule
				\multicolumn{15}{c}{\textit{Mean Square Forecast Error (MSFE)}}\\
				\midrule
				 $H$ & $\mathcal{M}_{1}$	&	$\mathcal{M}_{2}$	&	$\mathcal{M}_{3}$	&	$\mathcal{M}_{4}$	&	$\mathcal{M}_{5}$	&	$\mathcal{M}_{6}$	&	$\mathcal{M}_{7}$	&	$\mathcal{M}_{8}$	& $\mathcal{M}_{9}$  &
				$\mathcal{M}_{10}$   & $\mathcal{M}_{11}$ &
				$\mathcal{M}_{12}$ & $\mathcal{M}_{13}$	&	$\mathcal{M}_{14}$	&	$\mathcal{M}_{15}$	\\
				\midrule
				1	&	 0.5989 	&	 0.6178 	&	 0.5892 	&	 0.5890 	&	 0.7296 	&	 0.7154 	&	 0.6322 	&	 0.6429 	&	 0.8208 	&	 1.0254 	&	 --- 	&	 0.7875 	&	 1.0000 	&	 0.9621 	&	 0.8183 	\\
				2	&	 0.8072 	&	 0.8249 	&	 0.8071 	&	 0.8144 	&	 --- 	&	 --- 	&	 0.7610 	&	 0.7687 	&	 1.0886 	&	 1.6143 	&	 --- 	&	 1.0306 	&	 1.0000 	&	 0.6719 	&	 0.8359 	\\
				3	&	 0.8600 	&	 0.8550 	&	 0.8761 	&	 0.8743 	&	 --- 	&	 --- 	&	 0.9973 	&	 1.0003 	&	 1.4003 	&	 --- 	&	 --- 	&	 1.0625 	&	 1.0000 	&	 0.9011 	&	 1.2530 	\\
				4	&	 0.7182 	&	 0.7177 	&	 0.7308 	&	 0.7319 	&	 --- 	&	 --- 	&	 0.7338 	&	 0.7429 	&	 1.3066 	&	 --- 	&	 --- 	&	 0.7727 	&	 1.0000 	&	 0.8035 	&	 0.7715 	\\
				5	&	 1.0631 	&	 1.0512 	&	 1.0708 	&	 1.0689 	&	 --- 	&	 --- 	&	 1.1704 	&	 1.1830 	&	 --- 	&	 --- 	&	 --- 	&	 1.1073 	&	 1.0000 	&	 0.9719 	&	 1.1174 	\\
				6	&	 1.0669 	&	 1.0605 	&	 1.0625 	&	 1.0613 	&	 --- 	&	 --- 	&	 1.1639 	&	 1.1810 	&	 --- 	&	 --- 	&	 --- 	&	 1.0810 	&	 1.0000 	&	 0.9716 	&	 1.0828 	\\
				7	&	 1.0644 	&	 1.0581 	&	 1.0619 	&	 1.0610 	&	 --- 	&	 --- 	&	 1.0745 	&	 1.0781 	&	 --- 	&	 --- 	&	 --- 	&	 1.0817 	&	 1.0000 	&	 0.9738 	&	 1.0960 	\\
				8	&	 1.0607 	&	 1.0566 	&	 1.0535 	&	 1.0535 	&	 --- 	&	 --- 	&	 1.0864 	&	 1.0955 	&	 --- 	&	 --- 	&	 --- 	&	 1.0806 	&	 1.0000 	&	 0.9716 	&	 1.1063 	\\
				
				\midrule
				\multicolumn{15}{c}{\textit{Mean Absolute Forecast Error (MAFE)}}\\
				\midrule
				 H & $\mathcal{M}_{1}$	&	$\mathcal{M}_{2}$	&	$\mathcal{M}_{3}$	&	$\mathcal{M}_{4}$	&	$\mathcal{M}_{5}$	&	$\mathcal{M}_{6}$	&	$\mathcal{M}_{7}$	&	$\mathcal{M}_{8}$	& $\mathcal{M}_{9}$  &
				$\mathcal{M}_{10}$   & $\mathcal{M}_{11}$ &
				$\mathcal{M}_{12}$ & $\mathcal{M}_{13}$	&	$\mathcal{M}_{14}$	&	$\mathcal{M}_{15}$	\\
				\midrule
				1	&	 0.7548 	&	 0.7619 	&	 0.7570 	&	 0.7587 	&	 0.7605 	&	 0.7516 	&	 0.7452 	&	 0.7558 	&	 0.9040 	&	 0.7966 	&	 --- 	&	 0.7948 	&	 1.0000 	&	 0.9898 	&	 0.8505 	\\
				2	&	 0.8596 	&	 0.8707 	&	 0.8622 	&	 0.8648 	&	 0.9452 	&	 0.9485 	&	 0.8468 	&	 0.8514 	&	 0.9964 	&	 0.9609 	&	 --- 	&	 0.9345 	&	 1.0000 	&	 0.8203 	&	 0.9067 	\\
				3	&	 0.8783 	&	 0.8795 	&	 0.8910 	&	 0.8901 	&	 1.1748 	&	 1.1708 	&	 0.9266 	&	 0.9296 	&	 1.1909 	&	 1.2577 	&	 --- 	&	 0.9486 	&	 1.0000 	&	 0.9363 	&	 0.9600 	\\
				4	&	 0.8187 	&	 0.8199 	&	 0.8276 	&	 0.8283 	&	 1.3873 	&	 1.3912 	&	 0.8298 	&	 0.8321 	&	 1.1560 	&	 1.3600 	&	 --- 	&	 0.8291 	&	 1.0000 	&	 0.8865 	&	 0.8613 	\\
				5	&	 1.0076 	&	 1.0024 	&	 1.0085 	&	 1.0075 	&	 --- 	&	 --- 	&	 1.0399 	&	 1.0425 	&	 1.4274 	&	 --- 	&	 --- 	&	 1.0126 	&	 1.0000 	&	 0.9654 	&	 1.0453 	\\
				6	&	 1.0083 	&	 1.0051 	&	 1.0035 	&	 1.0035 	&	 --- 	&	 --- 	&	 1.0343 	&	 1.0363 	&	 1.4354 	&	 --- 	&	 --- 	&	 1.0063 	&	 1.0000 	&	 0.9629 	&	 1.0227 	\\
				7	&	 1.0048 	&	 1.0012 	&	 1.0019 	&	 1.0015 	&	 --- 	&	 --- 	&	 1.0121 	&	 1.0130 	&	 1.5180 	&	 --- 	&	 --- 	&	 1.0245 	&	 1.0000 	&	 0.9640 	&	 1.0284 	\\
				8	&	 0.9993 	&	 0.9957 	&	 0.9955 	&	 0.9953 	&	 --- 	&	 --- 	&	 0.9994 	&	 1.0042 	&	 1.5703 	&	 --- 	&	 --- 	&	 1.0126 	&	 1.0000 	&	 0.9589 	&	 1.0390 	\\
				\midrule
				\multicolumn{15}{c}{\textit{Average Log Predictive Likelihood (ALPL)}}\\
				\midrule
				\multicolumn{2}{c}{} & H & $\mathcal{M}_{1}$	&	$\mathcal{M}_{2}$	&	$\mathcal{M}_{3}$	&	$\mathcal{M}_{4}$	&	$\mathcal{M}_{5}$	&	$\mathcal{M}_{6}$	&	$\mathcal{M}_{7}$	&	$\mathcal{M}_{8}$	& $\mathcal{M}_{13}$	&	$\mathcal{M}_{14}$	&	$\mathcal{M}_{15}$	& \multicolumn{2}{c}{} \\
				\midrule
				\multicolumn{2}{c}{}	&	1	&	 1.7176 	&	 1.7277 	&	 1.6432 	&	 1.6455 	&	 1.3031 	&	 1.3031 	&	 1.3031 	&	 1.3031 	&	 1.0000 	&	 1.0046 	&	 1.4581 	&	\multicolumn{2}{c}{}	\\
				\multicolumn{2}{c}{}	&	2	&	 1.6248 	&	 1.6415 	&	 1.5459 	&	 1.5478 	&	 1.2565 	&	 1.2573 	&	 1.2575 	&	 1.2577 	&	 1.0000 	&	 0.9844 	&	 1.4621 	&	\multicolumn{2}{c}{}	\\
				\multicolumn{2}{c}{}	&	3	&	 1.5443 	&	 1.5535 	&	 1.4806 	&	 1.4803 	&	 1.2468 	&	 1.2477 	&	 1.2508 	&	 1.2515 	&	 1.0000 	&	 0.9807 	&	 1.4666 	&	\multicolumn{2}{c}{}	\\
				\multicolumn{2}{c}{}	&	4	&	 1.5332 	&	 1.5407 	&	 1.4712 	&	 1.4711 	&	 1.2321 	&	 1.2339 	&	 1.2368 	&	 1.2374 	&	 1.0000 	&	 0.9831 	&	 1.4566 	&	\multicolumn{2}{c}{}	\\
				\multicolumn{2}{c}{}	&	5	&	 1.5347 	&	 1.5430 	&	 1.4726 	&	 1.4725 	&	 1.2289 	&	 1.2306 	&	 1.2341 	&	 1.2349 	&	 1.0000 	&	 0.9821 	&	 1.4598 	&	\multicolumn{2}{c}{}	\\
				\multicolumn{2}{c}{}	&	6	&	 1.5264 	&	 1.5343 	&	 1.4634 	&	 1.4632 	&	 1.2206 	&	 1.2226 	&	 1.2253 	&	 1.2262 	&	 1.0000 	&	 0.9805 	&	 1.4576 	&	\multicolumn{2}{c}{}	\\
				\multicolumn{2}{c}{}	&	7	&	 1.5239 	&	 1.5319 	&	 1.4608 	&	 1.4606 	&	 1.2176 	&	 1.2197 	&	 1.2226 	&	 1.2235 	&	 1.0000 	&	 0.9798 	&	 1.4563 	&	\multicolumn{2}{c}{}	\\
				\multicolumn{2}{c}{}	&	8	&	 1.5143 	&	 1.5223 	&	 1.4514 	&	 1.4512 	&	 1.2087 	&	 1.2109 	&	 1.2139 	&	 1.2148 	&	 1.0000 	&	 0.9782 	&	 1.4474 	&	\multicolumn{2}{c}{}	\\
				\bottomrule
			\end{tabular}
	}}
\end{sidewaystable}

\begin{sidewaystable}
	\caption{Point and density forecast results for CPI Inflation. Root Mean Squared Forecast Error (RMSFE) upper panel, Median Absolute Forecast Error (MAFE),
		middle panel, Average Log Predictive Likelihood (ALPL) bottom panel. Results are reported relative to the benchmark specification
		($M_{13}$) for which the values is equal to 1, RMSFE-MAFE lower (higher) than 1 signify
		better (worse) performance than the benchmark. ALPL higher (lower) then 1 signify better (worse) performance. The description of the model is
		reported in Table \ref{tab:Multivariatemodels}. $---$ indicates divergence of the forecast.}%
	\label{tab:CPI}%
	\centering
	{\scriptsize {
			\begin{tabular}
				[c]{l|c|c|c|c|c|c|c|c|c|c|c|c|c|c|c}%
				\toprule
					\multicolumn{15}{c}{\textit{Mean Square Forecast Error (MSFE)}}\\
				\midrule
				$H$ & $\mathcal{M}_{1}$	&	$\mathcal{M}_{2}$	&	$\mathcal{M}_{3}$	&	$\mathcal{M}_{4}$	&	$\mathcal{M}_{5}$	&	$\mathcal{M}_{6}$	&	$\mathcal{M}_{7}$	&	$\mathcal{M}_{8}$	& $\mathcal{M}_{9}$  &
				$\mathcal{M}_{10}$   & $\mathcal{M}_{11}$ &
				$\mathcal{M}_{12}$ & $\mathcal{M}_{13}$	&	$\mathcal{M}_{14}$	&	$\mathcal{M}_{15}$	\\
				\midrule
			
				1	&	 1.0061 &  0.9992  &  1.0415 	&  1.0408  & 	 --- 	 & 	 --- 	 & 	 1.3206  & 	 1.3265 	 & 	 --- 	 & 	 1.6301  & 	 --- 	 & 	 0.8885 	 & 	 1.0000 	 & 	 0.9971 	 & 	 1.0621	\\
				2	&	 0.8491 &  0.8644  &  0.8311 	&  0.8330  & 	 --- 	 & 	 --- 	 & 	 0.9159  & 	 0.9197 	 & 	 --- 	 & 	 1.1349  & 	 --- 	 & 	 1.0564 	 & 	 1.0000 	 & 	 0.9398 	 & 	 0.7965	\\
				3	&	 0.8939 &  0.9081  &  0.8812 	&  0.8801  & 	 --- 	 & 	 --- 	 & 	 0.9126  & 	 0.9143 	 & 1.4230  & 	 1.4269  & 	 --- 	 & 	 0.8852 	 & 	 1.0000 	 & 	 0.9248 	 & 	 0.9016	\\
				4	&	 0.9667 &  0.9712  &  0.9639 	&  0.9648  & 	 --- 	 & 	 --- 	 & 	 1.0104  & 	 1.0107 	 & 	 --- 	 & 	 1.4161  & 	 --- 	 & 	 0.9749 	 & 	 1.0000 	 & 	 0.9557 	 & 	 0.9908	\\
				5	&	 1.0088 &  1.0114  &  1.0023 	&  1.0034  & 	 --- 	 & 	 --- 	 & 	 0.9809  & 	 0.9770 	 & 	 --- 	 & 	 --- 	   & 	 --- 	 & 	 1.0586 	 & 	 1.0000 	 & 	 0.9986 	 & 	 1.0296	\\
				6	&	 0.9899 &  0.9923  &  0.9909 	&  0.9908  & 	 --- 	 & 	 --- 	 & 	 1.0002  & 	 0.9998 	 & 	 --- 	 & 	 --- 	   & 	 --- 	 & 	 1.0180 	 & 	 1.0000 	 & 	 0.9995 	 & 	 1.0184	\\
				7	&	 0.9944 &  0.9941  &  0.9964 	&  0.9965  & 	 --- 	 & 	 --- 	 & 	 1.0099  & 	 1.0094 	 & 	 --- 	 & 	 --- 	   & 	 --- 	 & 	 1.0131 	 & 	 1.0000 	 & 	 0.9991 	 & 	 1.0029	\\
				8	&	 1.0010 &  0.9999  &  0.9995 	&  0.9996  & 	 --- 	 & 	 --- 	 & 	 1.0070  & 	 1.0066 	 & 	 --- 	 & 	 --- 	   & 	 --- 	 & 	 0.9936 	 & 	 1.0000 	 & 	 0.9989 	 & 	 1.0069	\\

				\midrule
				\multicolumn{15}{c}{\textit{Mean Absolute Forecast Error (MAFE)}}\\
				\midrule
					$H$ & $\mathcal{M}_{1}$	&	$\mathcal{M}_{2}$	&	$\mathcal{M}_{3}$	&	$\mathcal{M}_{4}$	&	$\mathcal{M}_{5}$	&	$\mathcal{M}_{6}$	&	$\mathcal{M}_{7}$	&	$\mathcal{M}_{8}$	& $\mathcal{M}_{9}$  &
				$\mathcal{M}_{10}$   & $\mathcal{M}_{11}$ &
				$\mathcal{M}_{12}$ & $\mathcal{M}_{13}$	&	$\mathcal{M}_{14}$	&	$\mathcal{M}_{15}$	\\
				\midrule
				1	&	 0.9731  & 	 0.9670  & 	 0.9942  & 	 0.9945  & 1.1331  &  1.1567 &  1.0499 &  1.0485 	 & 	 1.6029 	 & 	 1.0888 	 & 	 --- 	 & 	 0.9390 	 & 	 1.0000 	 & 	 0.9959 & 1.0039\\
				2	&	 0.9131  & 	 0.9233  & 	 0.9013  & 	 0.9027  & 1.1032  &  1.1091 &  0.9513 &  0.9564 	 & 	 1.4702 	 & 	 0.9717 	 & 	 --- 	 & 	 1.0166 	 & 	 1.0000 	 & 	 0.9722 & 0.8789\\
				3	&	 0.9201  & 	 0.9278  & 	 0.9128  & 	 0.9133  & 1.3179  &  1.3263 &  0.9349 &  0.9382 	 & 	 1.1703 	 & 	 1.0347 	 & 	 --- 	 & 	 0.9298 	 & 	 1.0000 	 & 	 0.9727 & 0.9459\\
				4	&	 0.9644  & 	 0.9675  & 	 0.9641  & 	 0.9642  & 1.8109  &  1.8308 &  1.0004 &  1.0019 	 & 	 1.4349 	 & 	 1.0776 	 & 	 --- 	 & 	 0.9780 	 & 	 1.0000 	 & 	 0.9781 & 0.9875\\
				5	&	 0.9985  & 	 0.9992  & 	 0.9939  & 	 0.9945  & 	 --- 	 & 	 --- 	 &  0.9813 &  0.9794 	 & 	 1.5745 	 & 	 1.2265 	 & 	 --- 	 & 	 1.0306 	 & 	 1.0000 	 & 	 0.9995 & 1.0213\\
				6	&	 0.9877  & 	 0.9899  & 	 0.9864  & 	 0.9864  & 	 --- 	 & 	 --- 	 &  0.9958 &  0.9954 	 & 	 1.3528 	 & 	 1.1740 	 & 	 --- 	 & 	 1.0076 	 & 	 1.0000 	 & 	 0.9969 & 1.0027\\
				7	&	 0.9886  & 	 0.9898  & 	 0.9903  & 	 0.9903  & 	 --- 	 & 	 --- 	 &  1.0076 &  1.0082 	 & 	 1.4098 	 & 	 1.5174 	 & 	 --- 	 & 	 1.0070 	 & 	 1.0000 	 & 	 0.9961 & 1.0055\\
				8	&	 0.9912  & 	 0.9904  & 	 0.9898  & 	 0.9898  & 	 --- 	 & 	 --- 	 &  0.9962 &  0.9965 	 & 	 1.6238 	 & 	 1.4867 	 & 	 --- 	 & 	 0.9883 	 & 	 1.0000 	 & 	 0.9937 & 0.9818\\

				\midrule
				\multicolumn{15}{c}{\textit{Average Log Predictive Likelihood (ALPL)}}\\
				\midrule
				\multicolumn{2}{c}{} & $H$ & $\mathcal{M}_{1}$	&	$\mathcal{M}_{2}$	&	$\mathcal{M}_{3}$	&	$\mathcal{M}_{4}$	&	$\mathcal{M}_{5}$	&	$\mathcal{M}_{6}$	&	$\mathcal{M}_{7}$	&	$\mathcal{M}_{8}$	& $\mathcal{M}_{13}$	&	$\mathcal{M}_{14}$	&	$\mathcal{M}_{15}$	& \multicolumn{2}{c}{} \\
				\midrule
				
				\multicolumn{2}{c}{}	&	1	&	 1.2543 	&	 1.2477 	&	 1.2441 	&	 1.2442 	&	 1.2681 	&	 1.2681 	&	 1.2681 	&	 1.2681 	&	 1.0000 	&	 1.0304 	&	 1.3009 	&	\multicolumn{2}{c}{}	\\
				\multicolumn{2}{c}{}	&	2	&	 1.1629 	&	 1.1525 	&	 1.1585 	&	 1.1553 	&	 1.2024 	&	 1.2002 	&	 1.1874 	&	 1.1871 	&	 1.0000 	&	 1.0189 	&	 1.3135 	&	\multicolumn{2}{c}{}	\\
				\multicolumn{2}{c}{}	&	3	&	 1.1704 	&	 1.1599 	&	 1.1652 	&	 1.1620 	&	 1.1837 	&	 1.1825 	&	 1.1846 	&	 1.1845 	&	 1.0000 	&	 1.0211 	&	 1.3233 	&	\multicolumn{2}{c}{}	\\
				\multicolumn{2}{c}{}	&	4	&	 1.1614 	&	 1.1475 	&	 1.1590 	&	 1.1555 	&	 1.1676 	&	 1.1669 	&	 1.1696 	&	 1.1696 	&	 1.0000 	&	 1.0155 	&	 1.3208 	&	\multicolumn{2}{c}{}	\\
				\multicolumn{2}{c}{}	&	5	&	 1.1446 	&	 1.1332 	&	 1.1441 	&	 1.1408 	&	 1.1342 	&	 1.1327 	&	 1.1363 	&	 1.1364 	&	 1.0000 	&	 1.0121 	&	 1.3106 	&	\multicolumn{2}{c}{}	\\
				\multicolumn{2}{c}{}	&	6	&	 1.1508 	&	 1.1398 	&	 1.1525 	&	 1.1491 	&	 1.1391 	&	 1.1372 	&	 1.1434 	&	 1.1435 	&	 1.0000 	&	 1.0124 	&	 1.3228 	&	\multicolumn{2}{c}{}	\\
				\multicolumn{2}{c}{}	&	7	&	 1.1443 	&	 1.1338 	&	 1.1469 	&	 1.1435 	&	 1.1321 	&	 1.1302 	&	 1.1372 	&	 1.1373 	&	 1.0000 	&	 1.0122 	&	 1.3175 	&	\multicolumn{2}{c}{}	\\
				\multicolumn{2}{c}{}	&	8	&	 1.1395 	&	 1.1291 	&	 1.1427 	&	 1.1393 	&	 1.1260 	&	 1.1240 	&	 1.1316 	&	 1.1317 	&	 1.0000 	&	 1.0121 	&	 1.3130 	&	\multicolumn{2}{c}{}	\\
				
				\bottomrule
			\end{tabular}
	}}
\end{sidewaystable}

\begin{sidewaystable}
	\caption{Point and density forecast results for Federal Funds Rate. Root Mean Squared Forecast Error (RMSFE) upper panel, Median Absolute Forecast Error (MAFE),
		middle panel, Average Log Predictive Likelihood (ALPL) bottom panel. Results are reported relative to the benchmark specification
		($M_{13}$) for which the values is equal to 1, RMSFE-MAFE lower (higher) than 1 signify
		better (worse) performance than the benchmark. ALPL higher (lower) then 1 signify better (worse) performance. The description of the model is
		reported in Table \ref{tab:Multivariatemodels}. $---$ indicates divergence of the forecast.}%
	\label{tab:FedFunds}%
	\centering
	{\scriptsize {
			\begin{tabular}
				[c]{l|c|c|c|c|c|c|c|c|c|c|c|c|c|c|c}%
				
				\toprule
					\multicolumn{14}{c}{\textit{Mean Square Forecast Error (MSFE)}}\\
				\midrule	
				$H$ & $\mathcal{M}_{1}$	&	$\mathcal{M}_{2}$	&	$\mathcal{M}_{3}$	&	$\mathcal{M}_{4}$	&	$\mathcal{M}_{5}$	&	$\mathcal{M}_{6}$	&	$\mathcal{M}_{7}$	&	$\mathcal{M}_{8}$	& $\mathcal{M}_{9}$  &
				$\mathcal{M}_{10}$   & $\mathcal{M}_{11}$ &
				$\mathcal{M}_{12}$ & $\mathcal{M}_{13}$	&	$\mathcal{M}_{14}$	&	$\mathcal{M}_{15}$	\\
				\midrule
				1	&	 1.0169 	&	 1.0492 	&	 0.9920 	&	 0.9975 	&	0.9432 &	0.8942 &	 1.2031 	&	 1.2534 &	 1.6167 	&	 1.4448	&	 --- 	&	 --- 	    &	 1.0000 	&	 1.0257 	&	 1.0797 	\\
				2	&	 1.0618 	&	 1.0669 	&	 1.0672 	&	 1.0686 	&	1.5283 &	1.5521 &	 1.2092 	&	 1.2049 &	 2.2307 	&	 1.6105	&	 --- 	&	 1.2776 	&	 1.0000 	&	 1.0070 	&	 1.1305 	\\
				3	&	 0.9844 	&	 0.9845 	&	 0.9880 	&	 0.9865 	&	 --- 	 &	 --- 	 &	 1.2535 	&	 1.2648 &	 1.8044 	&	 1.6272	&	 --- 	&	 1.2062 	&	 1.0000 	&	 1.0036 	&	 1.1549 	\\
				4	&	 1.0058 	&	 1.0084 	&	 1.0063 	&	 1.0103 	&	 --- 	 &	 --- 	 &	 1.0891 	&	 1.0907 &	 1.7889 	&	 --- 	  &	 --- 	&	 1.1566 	&	 1.0000 	&	 0.9999 	&	 1.2485 	\\
				5	&	 0.9807 	&	 0.9824 	&	 0.9878 	&	 0.9873 	&	 --- 	 &	 --- 	 &	 1.0565 	&	 1.0628 &	 1.6379 	&	 --- 	  &	 --- 	&	 1.1301 	&	 1.0000 	&	 0.9958 	&	 1.0943 	\\
				6	&	 0.9935 	&	 0.9957 	&	 0.9969 	&	 0.9967 	&	 --- 	 &	 --- 	 &	 1.0894 	&	 1.0908 &	 ---    	&	 --- 	  &	 --- 	&	 1.0798 	&	 1.0000 	&	 0.9901 	&	 1.1044 	\\
				7	&	 1.0107 	&	 1.0082 	&	 1.0066 	&	 1.0059 	&	 --- 	 &	 --- 	 &	 1.0662 	&	 1.0675 &	 --- 	    &	 --- 	  &	 --- 	&	 1.0645 	&	 1.0000 	&	 0.9906 	&	 0.9893 	\\
				8	&	 1.0005 	&	 1.0009 	&	 0.9973 	&	 0.9993 	&	 --- 	 &	 --- 	 &	 1.0577 	&	 1.0610 &	 --- 	    &	 --- 	  &	 --- 	&	 1.0392 	&	 1.0000 	&	 0.9927 	&	 1.0508 	\\

				\midrule
				\multicolumn{14}{c}{\textit{Mean Absolute Forecast Error (MAFE)}}\\
				\midrule
					$H$ & $\mathcal{M}_{1}$	&	$\mathcal{M}_{2}$	&	$\mathcal{M}_{3}$	&	$\mathcal{M}_{4}$	&	$\mathcal{M}_{5}$	&	$\mathcal{M}_{6}$	&	$\mathcal{M}_{7}$	&	$\mathcal{M}_{8}$	& $\mathcal{M}_{9}$  &
				$\mathcal{M}_{10}$   & $\mathcal{M}_{11}$ &
				$\mathcal{M}_{12}$ & $\mathcal{M}_{13}$	&	$\mathcal{M}_{14}$	&	$\mathcal{M}_{15}$	\\
				\midrule
				1	&	 0.9849 	&	 1.0078 	&	 0.9626 	&	 0.9679 	&	 1.1551 	&	 1.1323 &	 1.2447 	&	 1.2623 	&	 1.0871 	&	 1.2772 &	 --- 	&	 1.6658 	&	 1.0000 	&	 1.0376 	&	 1.0137 	\\
				2	&	 1.0021 	&	 1.0082 	&	 1.0000 	&	 1.0013 	&	 1.2102 	&	 1.2188 &	 1.2064 	&	 1.2076 	&	 1.4070 	&	 1.3413 &	 --- 	&	 1.2075 	&	 1.0000 	&	 1.0369 	&	 1.1817 	\\
				3	&	 0.9946 	&	 0.9957 	&	 0.9893 	&	 0.9890 	&	 1.3053 	&	 1.3146 &	 1.2017 	&	 1.2126 	&	 1.3457 	&	 1.2529 &	 --- 	&	 1.1237 	&	 1.0000 	&	 1.0325 	&	 1.0834 	\\
				4	&	 0.9819 	&	 0.9818 	&	 0.9783 	&	 0.9793 	&	 1.5724 	&	 1.5881 &	 1.1179 	&	 1.1264 	&	 1.3241 	&	 1.3810 &	 --- 	&	 1.0762 	&	 1.0000 	&	 1.0282 	&	 1.0904 	\\
				5	&	 0.9721 	&	 0.9724 	&	 0.9659 	&	 0.9652 	&	 1.9735 	&	 1.9804 &	 1.0762 	&	 1.0817 	&	 1.3743 	&	 1.3477 &	 --- 	&	 1.0754 	&	 1.0000 	&	 1.0220 	&	 1.0249 	\\
				6	&	 0.9789 	&	 0.9750 	&	 0.9715 	&	 0.9716 	&	 --- 	    &	 --- 	  &	 1.0784 	&	 1.0827 	&	 1.5579 	&	 1.5628 &	 --- 	&	 1.0099 	&	 1.0000 	&	 1.0202 	&	 1.0444 	\\
				7	&	 0.9795 	&	 0.9754 	&	 0.9718 	&	 0.9717 	&	 --- 	    &	 --- 	  &	 1.0756 	&	 1.0768 	&	 1.6735 	&	 1.6391 &	 --- 	&	 1.0108 	&	 1.0000 	&	 1.0189 	&	 0.9944 	\\
				8	&	 0.9715 	&	 0.9689 	&	 0.9672 	&	 0.9675 	&	 --- 	    &	 --- 	  &	 1.0400 	&	 1.0400 	&	 1.5535 	&	 --- 	  &	 --- 	&	 0.9869 	&	 1.0000 	&	 1.0182 	&	 1.0296 	\\

				\toprule
				
				\multicolumn{14}{c}{\textit{Average Log Predictive Likelihood (ALPL)}}\\
					\midrule
				\multicolumn{2}{c}{} & $H$ & $\mathcal{M}_{1}$	&	$\mathcal{M}_{2}$	&	$\mathcal{M}_{3}$	&	$\mathcal{M}_{4}$	&	$\mathcal{M}_{5}$	&	$\mathcal{M}_{6}$	&	$\mathcal{M}_{7}$	&	$\mathcal{M}_{8}$	& $\mathcal{M}_{13}$	&	$\mathcal{M}_{14}$	&	$\mathcal{M}_{15}$	& \multicolumn{2}{c}{} \\
			
				\midrule
				\multicolumn{2}{c}{}	&	1	&	1.4291	&	1.4162	&	1.3453	&	 1.3500 	&	 0.9299 	&	 0.9409 	&	0.9169	&	 0.9176 	&	 1.0000 	&	 1.0412 	&	1.0197	&	\multicolumn{2}{c}{}	\\
				\multicolumn{2}{c}{}	&	2	&	1.3384	&	1.3295	&	1.2789	&	 1.2839 	&	 0.9266 	&	 0.9297 	&	0.9159	&	 0.9177 	&	 1.0000 	&	 1.0100 	&	0.9865	&	\multicolumn{2}{c}{}	\\
				\multicolumn{2}{c}{}	&	3	&	1.2732	&	1.2668	&	1.2203	&	 1.2235 	&	 0.9336 	&	 0.9341 	&	0.9168	&	 0.9171 	&	 1.0000 	&	 1.0055 	&	1.0096	&	\multicolumn{2}{c}{}	\\
				\multicolumn{2}{c}{}	&	4	&	1.2434	&	1.2339	&	1.2038	&	 1.2063 	&	 0.9162 	&	 0.9166 	&	0.9153	&	 0.9167 	&	 1.0000 	&	 1.0038 	&	1.0127	&	\multicolumn{2}{c}{}	\\
				\multicolumn{2}{c}{}	&	5	&	1.2209	&	1.2116	&	1.1935	&	 1.1965 	&	 0.9207 	&	 0.9209 	&	0.9155	&	 0.9162 	&	 1.0000 	&	 1.0027 	&	1.0282	&	\multicolumn{2}{c}{}	\\
				\multicolumn{2}{c}{}	&	6	&	1.2157	&	1.2065	&	1.1896	&	 1.1920 	&	 0.9109 	&	 0.9061 	&	0.9145	&	 0.9160 	&	 1.0000 	&	 1.0035 	&	1.0208	&	\multicolumn{2}{c}{}	\\
				\multicolumn{2}{c}{}	&	7	&	1.2187	&	1.2072	&	1.1935	&	 1.1961 	&	 0.9170 	&	 0.9151 	&	0.9155	&	 0.9167 	&	 1.0000 	&	 1.0037 	&	1.0357	&	\multicolumn{2}{c}{}	\\
				\multicolumn{2}{c}{}	&	8	&	1.2271	&	1.2171	&	1.1989	&	 1.2013 	&	 0.9221 	&	 0.9190 	&	0.9235	&	 0.9249 	&	 1.0000 	&	 1.0032 	&	1.0415	&	\multicolumn{2}{c}{}	\\				
				
				\bottomrule
			\end{tabular}
	}}
\end{sidewaystable}
\noindent With three different variables, eight different forecast horizons
and two different forecast metrics, virtually every model can be found to do
well for some cases, but several observations can be made.\newline First,
TVP-MAI-SV is one of the best models for all the main variables, improving upon
its counterparts (MAI, TVP-MAI, MAI-SV) especially at the short horizon.\newline Second, an important
pattern emerges from the Tables. Adding SV to the MAI improve significantly 
the point and density forecast performances, this finding is in line with
\cite{CCM2019}. On all Tables the $\mathcal{M}_{7}$ and $\mathcal{M}_{8}$
models show worse performance than models $\mathcal{M}_{1}$, $\mathcal{M}_{2}%
$, $\mathcal{M}_{3}$ and $\mathcal{M}_{4}$. Moreover the $\mathcal{M}_{5}$ and
$\mathcal{M}_{6}$ show the poor performance of TVP-MAI homoskedastic
models.Tables \ref{tab:GDP}-\ref{tab:FedFunds} are a clear demonstration of
the importance of allowing for heteroskedastic errors to get good
RMSFE, MAFE and predictive likelihood.\newline One final point regards the
usefulness of adding time-varying parameters in the MAI-SV. The Tables show
that the two models have comparable point forecast performance, the TVP-MAI-SV
has always better predictive likelihood.\newline Overall, the results show that
the TVP-MAI-SV guarantees safe forecasting compared with the other
competitors such as the TVP-VAR-SV (see KK) and more similar models like the
TVP-FAVAR-SV as described in \cite{KK2013b}.

\section{Conclusions \label{sec:Concl}}

Many economic variables features changing mean and volatility, TVP-VAR with stochastic
volatility are commonly used to model those features. Starting from the recent MAI literature 
the paper introduces the TVP-MAI-SV that can handle large datasets. The paper 
introduces a new estimation methodology that substantially reduces the
computational burden, and allows to select in real time, the number of
indexes and other features of the data using DMS and DMA without further
computational cost.\\
The paper proposes two empirical applications.  The first provides a measure of 
uncertainty using the TVP-MAI-SV. We overcame endogeneity problems coming from the co-movement of macroeconomic variables by extracting a set of common unobservable factors representing the underlying aggregate 
uncertainty affecting the levels of the component variables. 
We get a rich set of volatility 
dynamics whose path provides interesting results in terms of common and idiosyncratic volatility changes.\\ 
The new switching algorithm, reduces the computational burden, allowing to apply to large databases (in our application we consider 215 series). From an economic point of view 
we were able to accurately capture both the Oil Crisis as well as the the Great Recession associated 
with much larger uncertainty shocks but no major changes in their effects on the economy.\\
The second empirical application shows the out-of-sample forecasting performance of the TVP-MAI-SV. Using both point and density forecast, we found that the TVP-MAI-SV model has good forecasting performance compared to a set 
of multivariate and univariate competitors.

\section*{Appendix A: Methodology} \label{sec:method} 
The model described in Section \ref{sec:Model}
allows for great flexibility e.g. different number of indexes over time.
From a technical point of view, handling such a large number of combinations is
not just computationally cumbersome but also memory intensive. \cite{RKE2010} and
\cite{KK2012} recently proposed the forgetting factor methodology that allows
for an online estimation of the time-varying parameters plus Dynamic Model
Averaging (DMA) and Dynamic Model Selection (DMS).\\
Following the discussion in Subsection \ref{subsec:Predictive}, let us consider $K$ 
possible models at each time point $t$. The number of models is a 
combination of number of indexes, 
values of $\lambda$ and values of $\kappa$. Considering all the possible
combination will increase exponentially the number of models.\\
The state space
model takes the following form:
\begin{equation}
\label{eq:SSM1}%
\begin{split}
\mathbf{y}_{t}  &=\mathbf{\mbox{Z}}_{t}^{(k)} \mbox{\boldmath $\beta$}_{t}%
^{(k)} +\mbox{\boldmath $\bvarepsilon$}_{t}^{(k)}, \hspace{1.5cm}
\mbox{\boldmath $\bvarepsilon$}_{t}^{(k)} \sim\N \left(  \boldsymbol{0}%
,\bH_{t}^{(k)}\right)  ,\\
\mbox{\boldmath $\beta$}_{t}^{(k)}  &=\mbox{\boldmath $\beta$}_{t-1}%
^{(k)}+\mbox{\boldmath $\eta$}_{t}^{(k)}, \hspace{2.0cm}
\mbox{\boldmath $\eta$}_{t}^{(k)}\sim\N \left(  \boldsymbol{0}%
,\Q^{(k)}\right)  ,
\end{split}
\end{equation}
where $\mathbf{y}_{t}$ is the vector of observed time series at time $t$, 
$\mathbf{Z}_{t}^{(k)}$ is the stack of
all the indexes depending on the unknown $\mbox{\boldmath $\omega$}$ (i.e. $\mathbf{Z}_{t}=I_N \otimes (f_{1,t},\ldots, f_{q,t})$ and $\f_{t} = \omega^{'}\by_t)$,
$\mbox{\boldmath $\beta$}_{t}^{(k)}$
is (possibly) an $Nq\times1$ vector containing the time varying 
$\mbox{\boldmath $\beta$}$s
(states), which are assumed to follow a random walk dynamic, see Subsection 
\ref{subsec:Bayes}. Finally $k = \{1, \ldots, K\}$ indicates the possible 
models based on a
specific ``sub"- set of: indexes, values of $\lambda$ and 
values of $\kappa$.\\	
The contemporaneous estimation of these models can be computationally cumbersome
and even be infeasible with maximum likelihood or MCMC methods. To overcome
this issue \cite{RKE2010} introduce an approximated KF that avoid the
calculation $\Q_{t}$ using a hyperparameter $\lambda$. \cite{KK2012}
applied this methodology to forecast inflation, they also add the estimation of
a time-varying $\H_{t}$ via EWMA that requires a decay factor $\kappa$.\\
The main step in the Kalman filter recursions, for a give model $k$, is given by:
\begin{equation}\label{eq:KF1}
\bbeta^{(k)}_{t-1}|\mbox{Y}_{t-1} \sim\N
\left(  \hat{\mbox{\boldmath $\beta$}}^{(k)}_{t-1|t-1}, \Sigma^{(k)}_{t-1|t-1}
\right)  ,
\end{equation}
where, $\mbox{Y}_{t-1}= \left(\mathbf{y}_{1},\mathbf{y}_{2}, \ldots
,\mathbf{y}_{t-1} \right)  $, $\hat{\mbox{\boldmath $\beta$}}^{(k)}%
_{t-1|t-1}=\mbox{E} \left(  \mbox{\boldmath $\beta$}^{(k)}_{t-1}%
|\mbox{Y}_{t-1} \right)  $ and $\Sigma^{(k)}_{t-1|t-1}=\mbox{Var} \left(
\mbox{\boldmath $\beta$}^{(k)}_{t-1}|\mbox{Y}_{t-1} \right)  $. At each time
point $t$, the algorithm iterates between: prediction equation, updating
equation and the predictive density:
\begin{equation}
\label{eq:KF2}\mbox{\boldmath $\beta$}^{(k)}_{t}|\mbox{Y}_{t-1} \sim\mbox{N}
\left(  \hat{\mbox{\boldmath $\beta$}}^{(k)}_{t|t-1}, \Sigma^{(k)}_{t|t-1}
\right)  ,
\end{equation}
\begin{equation}
\label{eq:KF3}\mbox{\boldmath $\beta$}^{(k)}_{t}|\mbox{Y}_{t} \sim\mbox{N}
\left(  \hat{\mbox{\boldmath $\beta$}}^{(k)}_{t|t}, \Sigma^{(k)}_{t|t}
\right)  ,
\end{equation}
\begin{equation}
\label{eq:KF4}\mathbf{y}_{t}|\mbox{Y}_{t-1} \sim\N \left(
\mathbf{\mbox{Z}}^{(k)}_{t} \hat{\mbox{\boldmath $\beta$}}^{(k)}_{t|t-1},
\bH^{(k)}_{t} + \mathbf{\mbox{Z}}^{(k)}_{t}\Sigma^{(k)}_{t|t-1}%
\mathbf{\mbox{Z}}^{(k)\prime}_{t} \right)  .
\end{equation}
The quantity $\Sigma_{t|t-1}^{(k)}$ depends on the error variances:
$\Sigma_{t|t-1}^{(k)}= \Sigma_{t-1|t-1}^{(k)} + \Q^{(k)}_{t}$.
\cite{RKE2010} proposed an approximation given by:
\begin{equation}
\label{eq:LAMBDA}\Sigma_{t|t-1}^{(k)}= \frac{1}{\lambda}\Sigma_{t-1|t-1}^{(k)}.%
\end{equation}
Correspondingly, $\Q^{(k)}_{t}=\left(  \frac{1}{\lambda} - 1 \right)
\Sigma^{(k)}_{t-1|t-1}$ with $\lambda\in(0,1]$. The tuning parameter $\lambda$
plays a crucial role in adjusting the effective memory of the algorithm,
leading to a weighted estimation where data at $i$ time points in the past has
weight $\lambda^{i}$. For example, in the case of quarterly 
macroeconomic data, $\lambda = 0.99$ implies that 
observations five years ago receive approximately 80\%
as much weight as the last period of observation. This leads to a fairly
stable model where coefficient change is gradual. When $\lambda = 1$, 
we have a constant parameter case. \\
It is well known that both macroeconomic and financial
time series are characterized by heteroskedastic effects, therefore, following
\cite{KK2012} $\bH^{(k)}_{t}$ is assumed to follow an EWMA:
\begin{equation}
\bH_{t} = \kappa\bH^{(k)}_{t-1}+\left(  1-\kappa\right) 
\bvarepsilon_{t}^{2 (k)}. \label{eq:riskmatrix}%
\end{equation}
where $\hat{\mbox{\boldmath $\varepsilon$}}_{t}^{(k)} = \mathbf{y}_{t} -
\mathbf{Z}_{t}\mbox{\boldmath $\beta$}_{t}$ is given as output from the Kalman
filter. The estimator that requires a value for the decay factor $\kappa$, 
following \cite{KK2013} the values for $\kappa$ are in 
the region of (0.94, 0.98). Moreover, the initial condition $\hat{\Sigma}_0$ is set to 
the sample covariance matrix of $\by_t$.\\
To carry out the model selection dynamically we use the following posterior
probabilities:
\begin{align}
\label{eq:PROB}
p\left(  \mbox{\boldmath $\beta$}_{t}, \mathcal{M}_{t}|\mbox{Y}_{t}
\right)   &  = \sum_{k=1}^{K}p \left(  \mbox{\boldmath $\beta$}_{t}%
^{(k)}|\mathcal{M}_{t}=k, \mbox{Y}_{t} \right)  p \left(  \mathcal{M}%
_{t}=k|\mbox{Y}_{t} \right)  =\nonumber\\
&  =\sum_{k=1}^{K}p \left(  \mbox{\boldmath $\beta$}_{t}^{(k)}%
|\mathcal{M}_{t}=k, \mbox{Y}_{t} \right)  \pi_{t|t,k}.
\end{align}
We need expression for the $p\left(  \mathcal{M}_{t}=
k|\mbox{Y}_{t} \right)$, similar to the state equation 
recursion, we have
the model prediction equation \eqref{eq:WEIGHT1} and the
the model
updating equation \eqref{eq:WEIGHT2}:
\begin{equation}
\label{eq:WEIGHT1}\pi_{t|t-1,k}=\sum_{l=1}^{K} \pi_{t-1|t-1,l}p_{kl},
\end{equation}
\begin{equation}
\label{eq:WEIGHT2}\pi_{t|t,k}=\frac{\pi_{t|t-1,k}p_{k} \left(  \mathbf{y}%
_{t}|\mbox{Y}_{t-1} \right)  }{\sum_{l=1}^{K} \pi_{t|t-1,l}p_{l} \left(
\mathbf{y}_{t}|Y_{t-1} \right)  }.
\end{equation}
where $p_{k} \left(\mathbf{y}_{t}|\Y_{t-1} \right)$ is the predictive density. We have to underline
that the model prediction equation requires to estimate the $K \times K$ elements of 
$p_{kl}$, this is replaced by an approximation:
\begin{equation}
\label{eq:WEIGHT1__}
\pi_{t|t-1,k}=\frac{\pi_{t-1|t-1,k}^{\alpha}}{\sum_{l=1}^{K} \pi_{t-1|t-1,l}^{\alpha}},
\end{equation} 
To interpret $\alpha$ let us take:
\begin{equation}
\label{eq:ALPHA}\pi_{t|t-1,k}\propto[\pi_{t-1|t-2,k}p_{k} \left(
\mathbf{y}_{t-1}|\mbox{Y}_{t-2} \right)  ]^{\alpha}=\prod_{i=1}^{t-1} [p_{k}
\left(  \mathbf{y}_{t-i}|\mbox{Y}_{t-i-1} \right)  ]^{\alpha^{i}}.
\end{equation}
where $p_{k} \left(  \mathbf{y}_{t-i}|\mbox{Y}_{t-i-1} \right)  $ is the
predictive density for the model $k$ evaluated at $\mathbf{y}_{t-i}$ with
$i=1,\ldots,t-1$.\newline The forgetting factor $\alpha\in(0,1]$ gives a
measure of the model performance rate of decay, the forecast performance
recorded $i$ periods in the past has a significance equal to $\alpha^{i}$.
Note that when $\alpha=0$ all models are equally probable for every $t$, the weights
of the models remain unchanged from the prior, $\pi_{0|0,k}=1/K$. Finally, 
\cite{KK2012} refer to the special case $\alpha=1$ as Bayesian Model Averaging
(BMA) which is very popular in macroeconomics and finance, see
\cite{KOOP_POTTER}.\\
From the recursive iteration a prediction for every model $k$ is obtained:
\begin{equation}
\label{eq:PRED}\mathbf{y}_{t}|\mathcal{M}_{t}=k, \mbox{Y}_{t-1} \sim\N
\left(  \mbox{Z}_{t}^{(k)} \hat{\mbox{\boldmath $\beta$}}_{t|t-1}^{(k)},
\bH_{t}^{(k)} + \mbox{Z}_{t}^{(k)}\Sigma_{t|t-1}^{(k)}\mbox{Z}_{t}%
^{(k)^{\prime}} \right)  .
\end{equation}
DMA comes from a weighted average of
all the the weights of the models that are the conditional 
probabilities
$P \left(  \mathcal{M}_{t}=k|\mbox{Y}_{t-1} \right)  =\pi_{t|t-1,k}$ computed
using the information up to time $t-1$ for $k=1,2, \ldots,K$:
\begin{equation}
\label{eq:DMAPRED}{\mathbf{y}_{DMA}}_{t}=\E \left(  \mathbf{y}_{t}%
|\mbox{Y}_{t-1} \right)  =\sum_{k=1}^{K}\pi_{t|t-1,k}\mbox{Z}_{t}^{(k)}
\hat{\mbox{\boldmath $\beta$}}_{t|t-1}^{(k)}.
\end{equation}
DMS selects and uses, at time $t$ the model with the highest 
predictive power capacity to make predictions of the dependent
variable.\\
The definition of a prior for $\pi_{0|0,k}$
and $\mbox{\boldmath $\beta$}_{0}^{(k)}$ is essential to implement DMA, DMS
and BMA. A non-informative prior is chosen for both the states and the
weights. In particular, $\pi_{0|0,k}={1}/{K}$ and
$\mbox{\boldmath $\beta$}_{0}^{(k)}\sim\N \left(  \boldsymbol{0}%
,\mbox{I} \right)  $ for $k=1,2,3, \ldots,K$. This means that, at the
beginning, all models are equally likely.

\section*{Appendix B:Factor Identification and Volatility Decomposition}

The structural analysis reported in Section \ref{sec:EmpApplication} requires
index identification. Following \cite{CCM2020} the structure of the
$\mbox{\boldmath $\omega$}$ matrix takes the following form:
\[
\mbox{\boldmath $\omega$}_{*}=\left[
\begin{array}
[c]{c|c|c|c|c}%
\mbox{RI} & \mbox{NI} & \mbox{LMI} & \mbox{PI} & \mbox{FI}\\
\midrule 1 & 0 & 0 & 0 & 0\\
\mbox{\boldmath $\omega$}_{2, RI} & 0 & 0 & 0 & 0\\
\vdots & \vdots & \ddots & \vdots & \vdots\\
\mbox{\boldmath $\omega$}_{n_{RI}, RI} & 0 & 0 & 0 & 0\\
0 & 1 & 0 & 0 & 0\\
0 & \mbox{\boldmath $\omega$}_{2, NI} & 0 & 0 & 0\\
\vdots & \vdots & \vdots & \vdots & \vdots\\
0 & \mbox{\boldmath $\omega$}_{n_{NI}, NI} & 0 & 0 & 0\\
0 & 0 & 1 & 0 & 0\\
0 & 0 & \mbox{\boldmath $\omega$}_{2, LMI} & 0 & 0\\
\vdots & \vdots & \vdots & \vdots & \vdots\\
0 & 0 & \mbox{\boldmath $\omega$}_{n_{LMI}, LMI} & 0 & 0\\
0 & 0 & 0 & 1 & 0\\
0 & 0 & 0 & \mbox{\boldmath $\omega$}_{1, PI} & 0\\
\vdots & \vdots & \vdots & \vdots & \vdots\\
0 & 0 & 0 & \mbox{\boldmath $\omega$}_{n_{PI}, PI} & 0\\
0 & 0 & 0 & 0 & 1\\
0 & 0 & 0 & 0 & \mbox{\boldmath $\omega$}_{2, FI}\\
\vdots & \vdots & \vdots & \vdots & \vdots\\
0 & 0 & 0 & 0 & \mbox{\boldmath $\omega$}_{n_{FI},FI}\\
&  &  &  &
\end{array}
\right]
\]
where $n = \{n_{RI}, n_{NI}, n_{LMI}, n_{PI}, n_{FI}\} $, the first series
load with a parameter equal to 1 and the rest are freely estimated. Please
note that we have $q = 5$ factors.\newline Tables \ref{t:fc1} to Table
\ref{t:fc5} contains the list of series that belong to each unknown factor. 
For example, Table \ref{t:fc1} report the series belonging to the first
factor, the GDP has loading equal to 1 and the rest are freely
estimated.\newline Equation \eqref{eq:Vec-model}, the OLS step of the SA
algorithm in Section \ref{sec:BayesEstimation}, can be re-written as:
\[
\underset{\mathbf{Y}_{\ast}}{\underbrace{(\H_{t}^{-1/2}\otimes
\mbox{I}_{T-p} )\mathrm{Vec}(\mathbf{Y})}}=\underset{\mathbf{X}_{\ast
}}{\underbrace{(\H_{t} ^{-1/2}\otimes\mbox{I}_{T-p})\mathbf{X}}%
}\mathrm{Vec}(\mbox{\boldmath $\omega$})+\mathrm{Vec}
(\mbox{\boldmath $\varepsilon$}_{t}\H_{t}^{-1/2}),
\]
where $\mathbf{Y} = [\mbox{Y}_{T}^{\prime},\ldots,\mbox{Y}_{p+1}^{\prime
}]^{\prime}$, $\mathbf{X}= {\textstyle\sum\limits_{j=1}^{p-1}} \beta
_{j,t}\otimes\mathbf{Y}_{-j}$, $\mathbf{Y}_{-j}=\mbox{L}^{j}\mathbf{Y}$, and
$\mbox{\boldmath $\varepsilon$}_{t}=(\varepsilon_{T}^{\prime},\ldots
\varepsilon_{p+1}^{\prime})^{\prime}$. Assuming that the matrix
$\mbox{\boldmath $\omega$}$\ has the following structure:
\[
\mbox{\boldmath $\omega$}=\left[
\begin{array}
[c]{cccc}%
\mbox{\boldmath $\omega$}_{1} & 0 & \cdots & 0\\
0 & \mbox{\boldmath $\omega$}_{2} & \cdots & 0\\
\vdots & \vdots & \ddots & \vdots\\
0 & \cdots & 0 & \mbox{\boldmath $\omega$}_{q}%
\end{array}
\right]  ,
\]
where $\omega_{i}$\ is an $n_{i}$-vector for $i=1,\ldots, q$ with $\sum
_{i=1}^{q}n_{i}=n$. Then $\mathrm{Vec}%
(\mbox{\boldmath $\omega$})=(\mbox{\boldmath $\omega$}_{1}^{\prime}%
,0_{n}^{\prime},\mbox{\boldmath $\omega$} _{2}^{\prime},0_{n}^{\prime}%
,\cdots,\mbox{\boldmath $\omega$}_{q}^{\prime})^{\prime}$ and can be factorized
as follows $\mathrm{Vec}%
(\mbox{\boldmath $\omega$})=\mbox{\boldmath $M$}\mbox{\boldmath $\omega$}_{\ast
}$ where $\mbox{\boldmath $\omega$}_{\ast}=\mathbf{(}%
\mbox{\boldmath $\omega$}_{1}^{\prime},\mbox{\boldmath $\omega$}_{2}^{\prime
},\cdots,\mbox{\boldmath $\omega$}_{q}^{\prime})^{\prime}$ and:
\[
\mbox{\boldmath $M$}=\left[
\begin{array}
[c]{cccc}%
\mbox{I}_{n_{1}} & 0_{n_{1}\times n_{2}} & \cdots & 0_{n_{1}\times n_{q}}\\
0_{n\times n_{1}} & 0_{n\times n_{2}} & \cdots & 0_{n\times n_{q}}\\
0_{n_{2}\times n1} & \mbox{I}_{n_{2}} & \cdots & 0_{n_{2}\times n_{q}}\\
0_{n\times n1} & 0_{n\times n_{2}} & \cdots & 0_{n\times n_{q}}\\
\vdots & \vdots & \ddots & \vdots\\
0_{n_{q}\times n1} & 0_{n_{q}\times n_{2}} & \cdots & \mbox{I}_{n_{q}}%
\end{array}
\right]  .
\]
It follows that the restricted OLS estimate of
$\mbox{\boldmath $\omega$}_{\ast}$ is:
\[
\hat{\mbox{\boldmath $\omega$}}_{\ast}=(\mbox{\boldmath $M$}^{\prime}
\mathbf{X}_{\ast}^{\prime}\mathbf{X}_{\ast}\mbox{\boldmath $M$})^{-1}%
\mathbf{X}_{\ast}^{\prime}\mbox{\boldmath $M$}^{\prime}\mathbf{Y}_{\ast}.
\vspace{-0.1cm}
\]
Following \cite{CCM2020} and \cite{CC2003} we have for the TVP-MAI-SV the
following decomposition:
\begin{equation}
\mbox{I}_{n} = \H_{t} \mbox{\boldmath $\omega$}_{*}^{^{\prime}}
\mbox{\boldmath $\xi$}_{t}^{-1} \mbox{\boldmath $\omega$}_{*}^{^{\prime}}+
\mbox{\boldmath $\omega$}_{*\perp}^{^{\prime}}
(\mbox{\boldmath $\omega$}_{*\perp}\H_{t}^{-1}
\mbox{\boldmath $\omega$}_{*\perp}^{^{\prime}})^{-1}
\mbox{\boldmath $\omega$}_{*\perp} \H_{t}^{-1},
\end{equation}
where $\mbox{\boldmath $\xi$}_{t} = \mbox{\boldmath $\omega$}_{*}
\H_{t} \mbox{\boldmath $\omega$}_{*}^{^{\prime}}$ and where
$\mbox{\boldmath $\omega$}_{*\perp}$ is the $(n-r) \times n$ orthogonal matrix
of $\mbox{\boldmath $\omega$}_{*}$, such that $\mbox{\boldmath $\omega$}_{*}%
\mbox{\boldmath $\omega$}_{*\perp}^{^{\prime}} = 0_{r \times(n-r)}$. Following
\cite{CCM2020} the total volatility $\H_{t}$ can be decomposed into the
volatility of the common and idiosyncratic components. Specifically:
\begin{equation}%
\begin{split}
\H_{t}  &  = \H^{com}_{t} + \H^{idio}_{t}\\
\H^{com}_{t}  &  = \H_{t} \mbox{\boldmath $\omega$}_{*}%
^{^{\prime}} \mbox{\boldmath $\xi$}_{t}^{-1} \mbox{\boldmath $\omega$}_{*}
\H_{t}\\
\H^{idio}_{t}  &  = \mbox{\boldmath $\omega$}_{*\perp}^{^{\prime}}
(\mbox{\boldmath $\omega$}_{*\perp} \H_{t}^{-1}
\mbox{\boldmath $\omega$}_{*\perp}^{^{\prime}})^{-1}%
\mbox{\boldmath $\omega$}_{*\perp}\\
\end{split}
\end{equation}

\section*{Appendix C: Dataset description}
\vspace{-0.3cm}
The quarterly data are download from Fred-Database and they
run from 1960:1 to 2019:4. All the
variables are transformed to achieve stationarity and then standardized using the Matlab code provided by FRED and available at the following web page  \url{https://research.stlouisfed.org/econ/mccracken/fred-databases/}, see \cite{MCNg2020}. To simplify the reading the full dataset, composed by 215 series, it is divided into five tables that represents the factor found and discussed in Section 
\ref{subsec:VarDec}. The series in bold are used in the forecasting exercise discussed in Section
\ref{subsec:Forecast}.

\begin{sidewaystable}
\caption{\textbf{Real Factor Composition}. The Table reports all the series used to extract the Real Factor divided in groups.
	All variables are transformed to be approximately
	stationary as in \cite{MCNg2020}. The table reports for each series the
	abbreviation, the full name of the series and the identifier
	in the FRED database. The Table also report, in bold, the series used in the forecasting exercise of Subsection \ref{subsec:Forecast}
}%
\label{t:fc1}%
\centering
{\tiny{
\begin{tabular}
[c]{r|l|l|l}%
\toprule N. Var & FRED-CODE & MEMO & Description	\\
\midrule
\multicolumn{4}{c}{\textit{Group 1: NIPA}}\\
\midrule
\textbf{1}  & \textbf{GDPC96} & \textbf{GDP} & \textbf{Real Gross Domestic Product, 3 Decimal (Billions of Chained 2012 Dollars)}\\
\textbf{2}  & \textbf{PCECC96} & \textbf{Consumption} & \textbf{Real Personal Consumption Expenditures (Billions of Chained 2012 Dollars)}\\
\textbf{3}  & \textbf{PINCOME} & \textbf{} & \textbf{Personal Income}\\
4  & PCDGx & Cons:Dur & Real personal consumption expenditures: Durable goods (Billions of Chained 2012 Dollars), deflated using PCE\\
5 & PCESVx & Cons:Svc & Real Personal Consumption Expenditures: Services (Billions of 2012 Dollars), deflated using PCE\\
6  & PCNDx & Cons:NonDur &  Real Personal Consumption Expenditures: Services (Billions of 2012 Dollars), deflated using PCE\\
7  & FPIx & FixedInv & Real private fixed investment (Billions of Chained 2012 Dollars), deflated using PCE\\
\textbf{8}  & \textbf{Y033RC1Q027SBEAx} & \textbf{GDPIC96} & \textbf{Software \& Real Gross Private Domestic Investment (Billions of Chained 2012 Dollars), deflated using PCE} \\
9  & PNFIx & FixInv:NonRes & Real private fixed investment: Nonresidential (Billions of Chained 2012 Dollars), deflated using PCE\\
10  & PRFIx & FixedInv:Res & Real private fixed investment: Residential (Billions of Chained 2012 Dollars), deflated using PCE\\
11 & A014RE1Q156NBEA & Inv:Inventories & Shares of gross domestic product: Gross private domestic investment: Change in private inventories (Percent)\\
12 & GCEC1& Gov.Spending & Real Government Consumption Expenditures \& Gross Investment (Billions of Chained 2012 Dollars)\\
13 & A823RL1Q225SBEA & Gov:Fed &  Real Government Consumption Expenditures and Gross Investment: Federal (Percent Change from Preceding Period)\\
14 & FGRECPTx & Real Gov Receipts & Real Federal Government Current Receipts (Billions of Chained 2012 Dollars), deflated using PCE \\
15 & SLCEx & Gov:State & Local \& Real government state and local consumption expenditures (Billions of Chained 2012 Dollars), deflated using PCE \\
16 & EXPGSC1 & Exports & Real Exports of Goods \& Services, 3 Decimal (Billions of Chained 2012 Dollars) \\
17 & IMPGSC1 & Imports & Real Imports of Goods \& Services, 3 Decimal (Billions of Chained 2012 Dollars) \\
18 & DPIC96 & Disp-Income & Real Disposable Personal Income (Billions of Chained 2012 Dollars) \\
19 & OUTNFB & Ouput:NFB & Nonfarm Business Sector: Real Output (Index 2012=100) \\
20 & OUTBS & Output:Bus &  Business Sector: Real Output (Index 2012=100) \\
21 & B020RE1Q156NBEA &  &Shares of gross domestic product: Exports of goods and services (Percent)\\
22 & B021RE1Q156NBEA &  & Shares of gross domestic product: Imports of goods and services (Percent) \\
\midrule
\multicolumn{4}{c}{\textit{Group 2: Industrial Production}}\\
\midrule
\textbf{23} & \textbf{INDPRO} & \textbf{IP:Total index} & \textbf{Industrial Production Index (Index 2012=100)}\\
24 & IPFINAL&IP:Final products & Industrial Production: Final Products (Market Group) (Index 2012=100)\\
25 & IPCONGD& IP:Consumer goods& Industrial Production: Consumer Goods (Index 2012=100)\\
26 & IPMAT& IP:Materials& Industrial Production: Materials (Index 2012=100)\\
27 & IPDMAT&IP:Dur gds materials & Industrial Production: Durable Materials (Index 2012=100)\\
28 & IPNMAT & IP:Nondur gds materials& Industrial Production: Nondurable Materials (Index 2012=100)\\
29 & IPDCONGD& IP:Dur Cons. Goods& Industrial Production: Durable Consumer Goods (Index 2012=100)\\
30 & IPB51110SQ& IP:Auto& Industrial Production: Durable Goods: Automotive products (Index 2012=100)\\
31 & IPNCONGD& IP:NonDur Cons God& Industrial Production: Nondurable Consumer Goods (Index 2012=100)\\
32 & IPBUSEQ& IP:Bus Equip& Industrial Production: Business Equipment (Index 2012=100)\\
33 & IPB51220SQ&IP:Energy Prds & Industrial Production: Consumer energy products (Index 2012=100)\\
\textbf{34} & \textbf{CUMFNS} & \textbf{Capu Man.}& \textbf{Capacity Utilization: Manufacturing (SIC) (Percent of Capacity)} \\
35 & IPMANSICS& &Industrial Production: Manufacturing (SIC) (Index 2012=100) \\
36 & IPB51222S& & Industrial Production: Residential Utilities (Index 2012=100)\\
37 & IPFUELS& & Industrial Production: Fuels (Index 2012=100)\\
\textbf{38} & & \textbf{PMI}& \textbf{Purchasing Managers Index}\\
\bottomrule
\end{tabular}
}}
\end{sidewaystable}

\begin{sidewaystable}
\caption{\textbf{Real Factor Composition continued}. The Table reports all the series used to extract the Real Factor divided in groups.
	All variables are transformed to be approximately
	stationary as in \cite{MCNg2020}. The table reports for each series the
	abbreviation we use in the paper, the full name of the series, the identifier
	in the FRED database. The Table also report in bold the series used in the forecasting exercise of Subsection \ref{subsec:Forecast}}%
\label{t:fc2}%
\centering
{\tiny{
\begin{tabular}
[c]{r|l|l|l}%
\toprule N. Var & FRED-CODE & MEMO & Description	\\
\midrule
\multicolumn{4}{c}{\textbf{\textit{Group 3: Earnings and Productivity}}}\\
\midrule
39 & CES3000000008x & Real AHE:MFG & Real Average Hourly Earnings of Production and Nonsupervisory Employees: Manufacturing (2012 Dollars per Hour), deflated by Core PCE\\
40 & COMPRNFB & CPH:NFB & Nonfarm Business Sector: Real Compensation Per Hour (Index 2012=100)\\
41 & RCPHBS & CPH:Bus & Business Sector: Real Compensation Per Hour (Index 2012=100)\\
42 & OPHNFB & OPH:nfb & Nonfarm Business Sector: Real Output Per Hour of All Persons (Index 2012=100)\\
\textbf{43} & \textbf{OPHBS} & \textbf{OPH:Bus} & \textbf{Business Sector: Real Output Per Hour of All Persons (Index 2012=100)}\\
44 & ULCBS & ULC:Bus & Business Sector: Unit Labor Cost (Index 2012=100)\\
45 & ULCNFB& ULC:NFB & Nonfarm Business Sector: Unit Labor Cost (Index 2012=100)\\
46 & UNLPNBS & UNLPay:nfb & Nonfarm Business Sector: Unit Nonlabor Payments (Index 2012=100)\\
47 & CES0600000008& & Average Hourly Earnings of Production and Nonsupervisory Employees: Goods-Producing (Dollars per Hour)\\
\midrule
\multicolumn{4}{c}{\textit{Group 4: Housing}}\\
\midrule
\textbf{48}& \textbf{HOUST} & \textbf{Hstarts} & \textbf{Housing Starts: Total: New Privately Owned Housing Units Started (Thousands of Units)}\\
49 & HOUST5F & Hstarts >5units & Privately Owned Housing Starts: 5-Unit Structures or More (Thousands of Units)\\
50 & HOUSTMW & Hstarts:MW & Housing Starts in Midwest Census Region (Thousands of Units)\\
51 & HOUSTNE & Hstarts:NE & Housing Starts in Northeast Census Region (Thousands of Units)\\
52 & HOUSTS & Hstarts:S & Housing Starts in South Census Region (Thousands of Units) \\
53 & HOUSTW & Hstarts:W & Housing Starts in West Census Region (Thousands of Units)\\
\midrule
\multicolumn{4}{c}{\textit{Group 5: Inventories, Orders and Sales}}\\
\midrule
54 & CMRMTSPLx & MT Sales & Real Manufacturing and Trade Industries Sales (Millions of Chained 2012 Dollars)\\
55 & RSAFSx & Ret. Sale & Real Retail and Food Services Sales (Millions of Chained 2012 Dollars), deflated by Core PCE\\
56 & AMDMNOx & Orders (DurMfg) & Real Manufacturers’ New Orders: Durable Goods (Millions of 2012 Dollars), deflated by Core PCE\\
57 & AMDMUOx & UnfOrders(DurGds) & Real Value of Manufacturers’ Unfilled Orders for Durable Goods Industries (Millions of 2012 Dollars), deflated by Core PCE\\
58 & BUSINVx & & Total Business Inventories (Millions of Dollars) \\
59 & ISRATIOx & & Total Business: Inventories to Sales Ratio\\
\textbf{60} & & \textbf{NAPMNOI} & \textbf{New Orders Manufacturing}\\
\bottomrule
\end{tabular}
}}
\end{sidewaystable}

\begin{sidewaystable}
\caption{\textbf{Nominal Factor Composition}. The Table reports all the series used to extract the Nominal Factor divided in groups.
	All variables are transformed to be approximately
	stationary as in \cite{MCNg2020}. The table reports for each series the
	abbreviation we use in the paper, the full name of the series, the identifier
	in the FRED database. The Table also report in bold the series used in the forecasting exercise of Subsection \ref{subsec:Forecast}}%
\label{t:fc3}%
\centering
{\tiny{
\begin{tabular}
[c]{r|l|l|l}%
\toprule N. Var & FRED-CODE & MEMO & Description	\\
\midrule
\multicolumn{4}{c}{\textit{Group 1: Interest Rates}}\\
\midrule
1 & TB3MS       & TB-3Mth  & 3-Month Treasury Bill: Secondary Market Rate (Percent)\\
2 & TB6MS       & TM-6MTH & 6-Month Treasury Bill: Secondary Market Rate (Percent)\\
3 & GS1         & TB-1YR & 1-Year Treasury Constant Maturity Rate (Percent)\\
4 & GS10        & TB-10YR & 10-Year Treasury Constant Maturity Rate (Percent)\\
5 & AAA         &AAA Bond & Moody’s Seasoned Aaa Corporate Bond Yield© (Percent)\\
6 & BAA         & BAA Bond& Moody’s Seasoned Baa Corporate Bond Yield© (Percent)\\
7 & BAA10YM     & BAA\_GS10 & Moody’s Seasoned Baa Corporate Bond Yield Relative to Yield on 10-Year Treasury Constant Maturity (Percent)\\
8 & TB6M3Mx     & tb6m\_tb3m& 6-Month Treasury Bill Minus 3-Month Treasury Bill, secondary market (Percent)\\
9 & GS1TB3Mx    & GS1\_tb3m & 1-Year Treasury Constant Maturity Minus 3-Month Treasury Bill, secondary market (Percent)\\
\textbf{10} & \textbf{GS10TB3Mx}  & \textbf{GS10\_tb3m} & \textbf{10-Year Treasury Constant Maturity Minus 3-Month Treasury Bill, secondary market (Percent)}\\
11 & CPF3MTB3Mx & CP\_Tbill Spread & 3-Month Commercial Paper Minus 3-Month Treasury Bill, secondary market (Percent)\\
12 & GS5        & & 5-Year Treasury Constant Maturity Rate\\
13 & TB3SMFFM   & & 3-Month Treasury Constant Maturity Minus Federal Funds Rate\\
14 & T5YFFM     & & 5-Year Treasury Constant Maturity Minus Federal Funds Rate\\
15 & AAAFFM     & & Moody’s Seasoned Aaa Corporate Bond Minus Federal Funds Rate\\
16 & CP3M       & & 3-Month AA Financial Commercial Paper Rate\\
17 & COMPAPFF   & & 3-Month Commercial Paper Minus Federal Funds Rate\\
\midrule
\multicolumn{4}{c}{\textit{Group 2: Exchange Rates}}\\
\midrule
\textbf{18} & \textbf{EXUSUKx} & \textbf{Ex rate:UK} & \textbf{U.S. / U.K. Foreign Exchange Rate}\\
19 & EXSZUSx & Ex rate:Switz & Switzerland / U.S. Foreign Exchange Rate\\
20 & EXJPUSx & Ex rate:Japan & Japan / U.S. Foreign Exchange Rate\\
21 & EXCAUSx & EX rate:Canada & Canada / U.S. Foreign Exchange Rate\\
\midrule
\multicolumn{4}{c}{\textit{Group 3: Money and Credit}}\\
\midrule
22 & BOGMBASEREALx & Real Mbase & Monetary Base (Millions of 1982-84 Dollars), deflated by CPI\\
23 & M1REAL & Real m1 & Real M1 Money Stock (Billions of 1982-84 Dollars), deflated by CPI\\
24 & M2REAL & Real m2 & Real M2 Money Stock (Billions of 1982-84 Dollars), deflated by CPI\\
25 & MZMREAL& Real mzm & Real MZM Money Stock (Billions of 1982-84 Dollars), deflated by CPI\\
26 & BUSLOANSx & Real C\&Lloand & Real Commercial and Industrial Loans, All Commercial Banks (Billions of 2012 U.S. Dollars), deflated by Core PCE\\
27 & CONSUMERx &Real ConsLoans & Real Consumer Loans at All Commercial Banks (Billions of 2012 U.S. Dollars), deflated by Core PCE\\
28 & NONREVSLx & Real NonRevCredit & Total Real Nonrevolving Credit Owned and Securitized, Outstanding (Billions of 2012 Dollars), deflated by Core PCE\\
29 & REALLNx & Real LoansRealEst & Real Real Estate Loans, All Commercial Banks (Billions of 2012 U.S. Dollars), deflated by Core PCE\\
30 & TOTALSLx & Real ConsuCred & Total Consumer Credit Outstanding (Billions of 2012 Dollars), deflated by Core PCE\\
31 & TOTRESNS & & Total Reserves of Depository Institutions (Billions of Dollars)\\
32 & NONBORRES & & Reserves Of Depository Institutions, Nonborrowed (Millions of Dollars) \\
33 & DTCOLNVHFNM & & Consumer Motor Vehicle Loans Outstanding Owned by Finance Companies (Millions of Dollars)\\
34 & DTCTHFNM & & Total Consumer Loans and Leases Outstanding Owned and Securitized by Finance Companies (Millions of Dollars)\\
35 & INVEST & & Securities in Bank Credit at All Commercial Banks (Billions of Dollars)\\
\textbf{36} & & \textbf{BORROW} & \textbf{Total Borrowings of Depository Institutions}\\
\textbf{37} & & \textbf{M1SL} & \textbf{M1 MOney Stock}\\
\textbf{38} & & \textbf{M2SL}  & \textbf{M2 Money Stock}\\
\bottomrule
\end{tabular}
}}
\end{sidewaystable}


\begin{sidewaystable}
\caption{\textbf{Labour Market Factor Composition}. The Table reports all the series used to extract the Labour Market Factor divided in groups.
	All variables are transformed to be approximately
	stationary as in \cite{MCNg2020}. The table reports for each series the
	abbreviation we use in the paper, the full name of the series, the identifier
	in the FRED database. The Table also report in bold the series used in the forecasting exercise of Subsection \ref{subsec:Forecast}}%
\label{t:fc4}%
\centering
{\tiny{\begin{tabular}[c]{r|l|l|l}%
\toprule N. Var & FRED-CODE & MEMO & Description	\\
\midrule
\multicolumn{4}{c}{\textit{Group 1: Employment and Unemployment}}\\
\midrule
\textbf{1} &  \textbf{PAYEMS}   & \textbf{Emp:Nonfarm} & \textbf{All Employees: Total nonfarm (Thousands of Persons)}\\
2 &  USPRIV   & Emp:Private & All Employees: Total Private Industries (Thousands of Persons)\\
3 &  MANEMP   & Emp:mfg & All Employees: Manufacturing (Thousands of Persons) \\
4 &  SRVPRD   & Emp:Services & All Employees: Service-Providing Industries (Thousands of Persons)\\
5 &  USGOOD   & Emp:Goods & All Employees: Goods-Producing Industries (Thousands of Persons)\\
6 &  DMANEMP   & Emp:DurGoods & All Employees: Durable goods (Thousands of Persons)\\
7 &  NDMANEMP  & Emp:Nondur Goods & All Employees: Nondurable goods (Thousands of Persons)\\
8 &  USCONS   & Emp:Const & All Employees: Construction (Thousands of Persons)\\
9 &  USEHS  &  Emp:Edu\&Health & All Employees: Education \& Health Services (Thousands of Persons)\\
10 & USFIRE   & Emp:Finance & All Employees: Financial Activities (Thousands of Persons)\\
11 & USINFO   & Emp:Infor & All Employees: Information Services (Thousands of Persons)\\
12 & USPBS   & Emp:Bus Serv & All Employees: Professional \& Business Services (Thousands of Persons)\\
13 & USLAH   & Emp:Leisure & All Employees: Leisure \& Hospitality (Thousands of Persons)\\
14 & USSERV   & Emp:OtherSvcs & All Employees: Other Services (Thousands of Persons)\\
15 &  USMINE   & Emp:Mining/NatRes & All Employees: Mining and logging (Thousands of Persons)\\
16 &  USTPU   & Emp:Trade\&Trans \& All Employees: Trade, Transportation & Utilities (Thousands of Persons)\\
17 &  USGOVT   &Emp:Gov  & All Employees: Government (Thousands of Persons)\\
18 &  USTRADE   & Emp:Retail & All Employees: Retail Trade (Thousands of Persons)\\
19 &  USWTRADE   & Emp:Wholesal & All Employees: Wholesale Trade (Thousands of Persons)\\
20 &  CES9091000001   & Emp:Gov(Fed) & All Employees: Government: Federal (Thousands of Persons) \\
21 &  CES9092000001   & Emp:Gov (State) & All Employees: Government: State Government (Thousands of Persons)\\
22 &  CES9093000001   &  Emp:Gov (Local) & All Employees: Government: Local Government (Thousands of Persons)\\
23 &  CE16OV   & Emp:Total (HHSurve) & Civilian Employment (Thousands of Persons)\\
24 &  CIVPART   &  LF Part Rate & Civilian Labor Force Participation Rate (Percent)\\
\textbf{25} &  \textbf{UNRATE}  & \textbf{Unemp Rate} & \textbf{Civilian Unemployment Rate (Percent)}\\
26 &  UNRATESTx   & Urate\_ST & Unemployment Rate less than 27 weeks (Percent)\\
27 &  UNRATELTx   & Urate\_LT & Unemployment Rate for more than 27 weeks (Percent)\\
28 &  LNS14000012   & Urate:Age16-19 & Unemployment Rate - 16 to 19 years (Percent)\\
29 &  LNS14000025   & Urate:Age>20 Men & Unemployment Rate - 20 years and over, Men (Percent)\\
30 &  LNS14000026   &  Urate:Age>20 Women & Unemployment Rate - 20 years and over, Women (Percent)\\
31 &  UEMPLT5   & U:Dur<5wks & Number of Civilians Unemployed - Less Than 5 Weeks (Thousands of Persons)\\
32 &  UEMP5TO14   & U:Dur5-14wks  & Number of Civilians Unemployed for 5 to 14 Weeks (Thousands of Persons)\\
33 &  UEMP15T26   & U:dur>15-26wks & Number of Civilians Unemployed for 15 to 26 Weeks (Thousands of Persons)\\
34 &  UEMP27OV   & U:Dur>27wks & Number of Civilians Unemployed for 27 Weeks and Over (Thousands of Persons)\\
35 &  LNS12032194   & Emp:SlackWk & Employment Level - Part-Time for Economic Reasons, All Industries (Thousands of Persons)\\
36 &  HOABS   & EmpHrs:Bus Sec & Business Sector: Hours of All Persons (Index 2012=100)\\
37 &  HOANBS    & EmpHrs:nfb & Nonfarm Business Sector: Hours of All Persons (Index 2012=100)\\
\textbf{38} &                & \textbf{AHMAN} & \textbf{Average Hourly Manufacturing}\\
39 &  AWHMAN   & AWH Man & Average Weekly Hours of Production and Nonsupervisory Employees: Manufacturing (Hours)\\
40 &  AWOTMAN   & AWH Overtime & Average Weekly Overtime Hours of Poduction and Nonsupervisory Employees: Manufacturing (Hours)\\
41 &  HWIx    & HelpWnted & Help-Wanted Index\\
42 &  UEMPMEAN   &  & Average (Mean) Duration of Unemployment (Weeks)\\
43 &  CES0600000007   &  & Average Weekly Hours of Production and Nonsupervisory Employees: Goods-Producing\\
44 &  HWIURATIOx   &  & Ratio of Help Wanted/No. Unemployed\\
45 &  CLAIMSx   &  & Initial Claims\\
\bottomrule
\end{tabular}
}}
\end{sidewaystable}

\begin{sidewaystable}
\caption{\textbf{Prices Factor Composition}. The Table reports all the series used to extract the Prices Factor divided in groups.
	All variables are transformed to be approximately
	stationary as in \cite{MCNg2020}. The table reports for each series the
	abbreviation we use in the paper, the full name of the series, the identifier
	in the FRED database. The Table also report in bold the series used in the forecasting exercise of Subsection \ref{subsec:Forecast}}%
\label{t:fc5}%
\centering
{\tiny{\begin{tabular}
[c]{r|l|l|l}%
\toprule N. Var & FRED-CODE & MEMO & Description	\\
\midrule
\multicolumn{4}{c}{\textit{Group 1: Prices}}\\
\midrule
\textbf{1} &  \textbf{CPIAUCSL}   & \textbf{CPI} & \textbf{Consumer Price Index for All Urban Consumers: All Items (Index 1982-84=100)}\\
\textbf{2} &  \textbf{PCECTPI}   & \textbf{PCED} & \textbf{Personal Consumption Expenditures: Chain-type Price Index (Index 2012=100)}\\
\textbf{3} &  \textbf{PPIACO}   &  \textbf{PPI} & \textbf{Producer Price Index for All Commodities (Index 1982=100)}\\
4 &  PCEPILFE   &  PCED\_LFE & Personal Consumption Expenditures Excluding Food and Energy (Chain-Type Price Index) (Index 2012=100)\\
5 &  GDPCTPI  & GDP Defl & Gross Domestic Product: Chain-type Price Index (Index 2012=100)\\
6 &  GPDICTPI   & GPDI Defl & Gross Private Domestic Investment: Chain-type Price Index (Index 2012=100)\\
7 &  IPDBS  & BusSec Defl & Business Sector: Implicit Price Deflator (Index 2012=100)\\
8 &  DGDSRG3Q086SBEA   & PCED\_Goods & Personal consumption expenditures: Goods (chain-type price index)\\
9 &  DDURRG3Q086SBEA   & PCED\_Serv & Personal consumption expenditures: Services (chain-type price index) \\
10 & DSERRG3Q086SBEA    & PCED\_DurGoods & Personal consumption expenditures: Durable goods (chain-type price index)\\
11 & DNDGRG3Q086SBEA    & PCED\_NDurGoods & Personal consumption expenditures: Nondurable goods (chain-type price index)\\
12 & DHCERG3Q086SBEA    & PCED\_HouseholdServ & Personal consumption expenditures: Services: Household consumption expenditures (chain-type price index)\\
13 & DMOTRG3Q086SBEA    & CED\_MotorVec  & Personal consumption expenditures: Durable goods: Motor vehicles and parts (chain-type price index)\\
14 & DFDHRG3Q086SBEA    & PCED\_DurHousehold & Personal consumption expenditures: Furnishings and durable household equipment (chain-type price index)\\
15 & DREQRG3Q086SBEA    &  PCED\_Recreation & Personal consumption expenditures: Durable goods: Recreational goods
and vehicles (chain-type price index)\\
16 & DODGRG3Q086SBEA    & PCED\_OthDurGds & Personal consumption expenditures: Durable goods: Other durable goods
(chain-type price index)\\
17 & DFXARG3Q086SBEA    & PCED\_Food\_Bev & Personal consumption expenditures: Food and beverages purchased for off-premises consumption (chain-type price index)\\
18 & DCLORG3Q086SBEA    & PCED\_Clothing & Personal consumption expenditures: Nondurable goods: Clothing and
footwear (chain-type price index)\\
19 &  DGOERG3Q086SBEA   & PCED\_Gas\_Enrgy & Personal consumption expenditures: Nondurable goods: Gasoline and other energy goods (chain-type price index)\\
20 &  DONGRG3Q086SBEA	  & PCED\_OthNDurGds & Personal consumption expenditures: Nondurable goods: Other nondurable goods (chain-type price index)\\
21 &  DHUTRG3Q086SBEA   & PCED\_Housing-Utilities & Personal consumption expenditures: Services: Housing and utilities (chain-type price index)\\
22 &  DHLCRG3Q086SBEA   & PCED\_HealthCare & Personal consumption expenditures: Services: Health care (chain-type price index) \\
23 &  DTRSRG3Q086SBEA   & PCED\_TransSvg & Personal consumption expenditures: Transportation services (chain-type price index)\\
24 &  DRCARG3Q086SBEA   & PCED\_RecServices & Personal consumption expenditures: Recreation services (chain-type price index)\\
25 &  DFSARG3Q086SBEA   & PCED\_FoodServ\_Acc & Personal consumption expenditures: Services: Food services and accommodations (chain-type price index)\\
26 &  DIFSRG3Q086SBEA   & PCED\_FIRE & Personal consumption expenditures: Financial services and insurance (chain-type price index)\\
27 &  DOTSRG3Q086SBEA	& PCED\_OtherServices & Personal consumption expenditures: Other services (chain-type price index)\\
28 &  CPILFESL   & CPI\_LFE & Consumer Price Index for All Urban Consumers: All Items Less Food \& Energy (Index 1982-84=100)\\
29 &  WPSFD49502   & PPI:FinConsGds  & Producer Price Index by Commodity for Finished Consumer Goods (Index 1982=100)\\
30 &  WPSFD4111   & PPI:FinConsGds(Food) & Producer Price Index by Commodity for Finished Consumer Foods (Index 1982=100)\\
31 &  PPIIDC   & PPI:IndCom & Producer Price Index by Commodity Industrial Commodities (Index 1982=100)\\
32 &  WPSID61    & PPI:IntMat & Producer Price Index by Commodity Intermediate Materials: Supplies \& Components (Index 1982=100)\\
33 &  WPU0561    & Real Price:Oil & Producer Price Index by Commodity for Fuels and Related Products and Powe (Index 1982=100) \\
\textbf{34} &  \textbf{OILPRICEx}   & \textbf{Real Crudeoil Price} & \textbf{Real Crude Oil Prices: West Texas Intermediate (WTI) - Cushing, Oklahoma (2012 Dollars per Barrel), deflated by Core PCE}\\
35 &   WPSID62  &  & Producer Price Index: Crude Materials for Further Processing (Index 1982=100)\\
36 &   PPICMM  &  & Producer Price Index: Commodities: Metals and metal products: Primary nonferrous metals (Index 1982=100)\\
37 &   CPIAPPSL  &  & Consumer Price Index for All Urban Consumers: Apparel (Index
1982-84=100)\\
38 &   CPITRNSL  &  & Consumer Price Index for All Urban Consumers: Transportation (Index 1982-84=100) \\
39 &   CPIMEDSL  &  & Consumer Price Index for All Urban Consumers: Medical Care (Index 1982-84=100)\\
40 &   CUSR0000SAC  &  & Consumer Price Index for All Urban Consumers: Commodities (Index 1982-84=100) \\
41 &   CUSR0000SAD  &  & Consumer Price Index for All Urban Consumers: Durables (Index 1982-84=100)\\
42 &   CUSR0000SAS  &  & Consumer Price Index for All Urban Consumers: Services (Index 1982-84=100)\\
43 &   CPIULFSL  &  & Consumer Price Index for All Urban Consumers: All Items Less Food (Index 1982-84=100)\\
44 &   CUSR0000SA0L2  &  & Consumer Price Index for All Urban Consumers: All items less shelter (Index 1982-84=100)\\
45 &   CUSR0000SA0L5  &  & Consumer Price Index for All Urban Consumers: All items less medical care (Index 1982-84=100)\\
\textbf{46} &   \textbf{NAPMPRI}  &  & \textbf{CPI for All Urban Consumers: Owners’ equivalent rent of residences (Index Dec 1982=100)}\\
\bottomrule
\end{tabular}
}}
\end{sidewaystable}

\begin{sidewaystable}
\caption{Financial Factor Composition. The Table reports all the series used to extract the Financial Factor divided in groups.
	All variables are transformed to be approximately
	stationary as in \cite{MCNg2020}. The table reports for each series the
	abbreviation we use in the paper, the full name of the series, the identifier
	in the FRED database. The Table also report in bold the series used in the forecasting exercise of Subsection \ref{subsec:Forecast}}%
\label{t:fc6}%
\centering
{\tiny{\begin{tabular}
[c]{r|l|l|l}%
\toprule N. Var & FRED-CODE & MEMO & Description	\\
\midrule
\multicolumn{4}{c}{\textit{Group 1: Stock Markets}}\\
\midrule
\textbf{1}  &  \textbf{FEDFUNDS}         & \textbf{FedFunds} & \textbf{Effective Federal Funds Rate (Percent)}\\
2  &  NIKKEI225        &  & Nikkei Stock Average\\
3  &  S\&P:indust       &  & S\&P's Common Stock Price Index: Industrials\\
4  &  S\&P div yield    &  & S\&P's Composite Common Stock: Dividend Yield\\
5  &  S\&P PE ratio     &  & S\&P's Composite Common Stock: Price-Earnings Ratio\\
\textbf{6}  &  \textbf{SP500}            &  & \textbf{S\&P's Common Stock Price Index: Composite}\\
\midrule
\multicolumn{4}{c}{\textit{Group 2: Household Balance Sheet}}\\
\midrule
7  &  TABSHNOx         & Real HHW:TASA & Real Total Assets of Households and Nonprofit Organizations (Billions of 2012 Dollars), deflated by Core PCE\\
8  &  TLBSHNOx         & Real HHW:LiabSA & Real Total Liabilities of Households and Nonprofit Organizations (Billions of 2012 Dollars), deflated by Core PCE\\
9  &  LIABPIx          & liab\_PDISA & Liabilities of Households and Nonprofit Organizations Relative to Personal Disposable Income (Percent) \\
10 &  TNWBSHNOx        & Real HHW:WSA  & Real Net Worth of Households and Nonprofit Organizations (Billions of 2012 Dollars), deflated by Core PCE\\
11 &  NWPIx            & W\_PDISA & Net Worth of Households and Nonprofit Organizations Relative to Disposable Personal Income (Percent)\\
12 &  TARESAx          & Real HHW:TA\_RESA & Real Assets of Households and Nonprofit Organizations excluding Real Estate Assets\\
&                   &                   & (Billions of 2012 Dollars), deflated by Core PCE\\
13 &  HNOREMQ027Sx     & Real HHW:RESA  & Real Real Estate Assets of Households and Nonprofit Organizations (Billions of 2012 Dollars), deflated by Core PCE\\
14 &  CONSPIx          &  & Nonrevolving consumer credit to Personal Income\\
15 &  TFAABSHNOx       & Real HHW:FinSA & Real Total Financial Assets of Households and Nonprofit Organizations (Billions of 2012 Dollars), deflated by Core PCE\\
\midrule
\multicolumn{4}{c}{\textit{Group 3: Non-Household Balance Sheet}}\\
\midrule
16 &  TLBSNNCBx        &  & Real Nonfinancial Corporate Business Sector Liabilities (Billions of 2012 Dollars), Deflated by IPDBS\\
17 &  TLBSNNCBBDIx     &  & Nonfinancial Corporate Business Sector Liabilities to Disposable Business Income (Percent)\\
18 &  TTAABSNNCBx      &  & Real Nonfinancial Corporate Business Sector Assets (Billions of 2012 Dollars), Deflated by IPDBS\\
19 &  TNWMVBSNNCBx     &  & Real Nonfinancial Corporate Business Sector Assets (Billions of 2012 Dollars), Deflated IPDBS \\
20 &  TNWMVBSNNCBBDIx  &  & Nonfinancial Corporate Business Sector Net Worth to Disposable Business Income (Percent)\\
21 &  TLBSNNBx         &  & Real Nonfinancial Noncorporate Business Sector Liabilities (Billions of 2012 Dollars), Deflated by IPDBS\\
22 &  TLBSNNBBDIx      &  & Nonfinancial Noncorporate Business Sector Liabilities to Disposable Business Income (Percent)\\
23 &  TABSNNBx         &  & Real Nonfinancial Noncorporate Business Sector Assets (Billions of 2012 Dollars), Deflated by IPDBS\\
24 &  TNWBSNNBx        &  & Real Nonfinancial Noncorporate Business Sector Net Worth (Billions of 2012 Dollars), Deflated by IPDBS\\
25 &  TNWBSNNBBDIx     &  & Nonfinancial Noncorporate Business Sector Net Worth to Disposable Business Income (Percent)\\
26 &  CNCFx            &  & Real Disposable Business Income, Billions of 2012 Dollars (Corporate cash flow with IVA minus taxes\\
&                   &  & on corporate income, deflated by Implicit Price Deflator for Business Sector IPDBS)\\
\bottomrule
\end{tabular}
}}
\end{sidewaystable}


\clearpage
\singlespacing
\bibliographystyle{apalike}
\bibliography{Biblio}

\end{document}